\documentclass[twocolumn]{aastex631}
\usepackage{amsmath}
\usepackage{graphicx}
\usepackage{subfigure}
\usepackage{units}
\usepackage{natbib}
\usepackage{comment}

\graphicspath{{./}{figures/}}

\begin{document}

\title{UNIPOLARITY OF THE SOLAR MAGNETIC FIELD IN EQUATORIAL CORONAL HOLES}

\author{Khagendra Katuwal}
\affiliation{Department of Astronomy, New Mexico State University, Las Cruces, NM 88003, USA}

\author{R.T. James McAteer}
\affiliation{Department of Astronomy, New Mexico State University, Las Cruces, NM 88003, USA}

\begin{abstract}
A study of the unbalanced magnetic polarity distribution of 70 coronal holes was performed. Data from the Helioseismic and Magnetic Imager (HMI) were used to examine the photospheric line-of-sight magnetic field ($B_{\mathrm{LOS}}$) beneath these coronal holes. The skewness ($S$) values of the $B_{\mathrm{LOS}}$ distributions revealed significant asymmetry, characterized by the dominance of one magnetic polarity, with $\sim88\%$ of the coronal holes exhibiting a skewness value ranging from $\pm(0.20~\text{to}~0.40)$. The corresponding magnetic flux imbalance ($\Phi_{\mathrm{imb}}$) ranges from $20\%$ to $45\%$. In contrast, quiet-Sun regions show symmetric magnetic field distributions with skewness values less than$~0.11$ and flux imbalance less than $11.0\%$. A study of a coronal hole as it traverses across the disk shows that the magnetic field distribution does not evolve significantly over this time, remaining stable across half a solar rotation. A moderate correlation ($r = 0.60$) between the magnetic flux imbalance and the speed of associated high speed solar wind streams ($v_{\mathrm{HSS}}$) suggests that flux imbalance may contribute to the generation of these faster solar wind streams. These results imply that regions with higher flux imbalance ($\Phi_{\mathrm{imb}}$), indicative of more open magnetic field structures, present more favorable conditions for plasma acceleration as compared to closed bi-polar field, but the moderate correlation indicates that other factors may also play important roles.
\end{abstract}

\section{Introduction} \label{sec:intro}
The Sun's magnetic field underlies all forms of solar features and activity, including coronal holes, solar flares, and coronal mass ejections. For coronal holes, studying the distribution of the magnetic field lines in the photosphere enables us to better understand the acceleration of solar plasma to solar wind speeds in the range of around 250-800 $\,\mathrm{km\,s^{-1}}$. This, in turn, improves forecasting accuracy and readiness for potential impacts on Earth and space-based technologies. High speed solar wind streams originating from coronal holes can interact with the preceding slower solar wind, forming co-rotating interaction regions (CIRs). These CIRs compress plasma and magnetic fields, often leading to geomagnetic storms, auroral enhancements, and disturbances in satellites, communications, and power systems. Hence, understanding the magnetic distribution of coronal holes is crucial for both comprehending solar activity and predicting space weather more precisely.

Coronal holes are considered to be the sources of the fast solar wind \citep{1973SoPh...29..505K,1976SoPh...46..303N,wang1990solar,mccomas2007understanding}. However, the specific energy source and acceleration mechanisms of such highly energized solar plasma originating from coronal holes are still an open question. Coronal holes are temporary regions which appear dark in images of the solar corona, most commonly in the extreme ultraviolet (EUV) and soft X-ray \citep{CRANMER20023,2022ApJ...932..115S}. They exhibit low density, temperature, and intensity emission compared to non-coronal hole regions \citep{1972ApJ...176..511M}.
The magnetic field within coronal holes is considered to be open \citep{altschuler1972coronal} and unipolar, originating at or below the photosphere and extending outward into interplanetary space \citep{cranmer2009coronal}. These open magnetic field lines play an important role in the formation of the fast solar wind. Open magnetic fields easily allow plasma to escape with velocities around $400-800$ $\,\mathrm{km\,s^{-1}}$ \citep{2006SSRv..124...51S} from the corona, whereas closed magnetic fields trap plasma, resulting in slower speeds \citep{2004ApJ...612.1171W,2006SSRv..124...51S}.

Although they tend to follow a solar cycle variation, coronal holes can form at any location and time during the 22-year solar cycle. In addition to changing shape and size as the Sun's activity cycle progresses from minimum to maximum, they also tend to show a preferred hemispheric polarity. The study of the photospheric magnetic field distribution in coronal holes, before formation, during their period of relative stability, and after their decay is crucial for understanding the exact connection between the magnetic environment of coronal holes and the global magnetic field structure of the Sun.

Coronal holes were first observed as dark patches during solar eclipses from 1950 to 1960, although their nature was not understood at that time. The Skylab space station, launched in 1973, carried X-ray telescopes that provided the first comprehensive images of coronal holes \citep{1974ApJ...194L.115H,1973SoPh...29..505K}. The study of coronal holes has made great progress in the last few decades. Advancements in theoretical understanding, computational modeling, observation techniques, different spacecraft viewpoints, and ground-based instruments have significantly improved our understanding of coronal hole magnetic fields.

Coronal holes are now understood to be magnetically driven phenomena, with their plasma properties, such as temperature and density, being closely linked to the magnetic field configuration on the Sun. Direct and accurate measurements of the magnetic field in the solar corona are difficult, because of the weak magnetic field strength, high temperature and low densities \citep{Kucera2002}. The world's most powerful solar telescope, the Daniel K. Inouye Solar Telescope (DKIST), will provide the highest-resolution data at various heights of the solar atmosphere for small portions of coronal holes. At the photospheric height, the magnetic fields are strong enough to cause significant splitting of spectral lines (Zeeman effect), which can be measured and analyzed to determine both the strength and orientation of the magnetic field \citep{Kucera2002}. Simulations suggest that individual photospheric magnetic structures may be as small as 0.05\arcsec, equivalent to 35 km \citep{2003ESASP.530..279S}.

Here, we use two parameters to quantify the unipolar degree of the magnetic field distribution of coronal holes in the photosphere - the flux imbalance ($\Phi_{\mathrm{\textbf{imb}}}$) and the skewness ($S$) - and  probe the relation of this flux imbalance to high speed solar wind streams. We define unipolar as the dominance of one magnetic polarity, either positive magnetic flux (field lines considered by convention as emanating away from the sun) or negative magnetic flux (field lines considered by convention as pointing towards the sun). This degree of magnetic unipolarity is a key characteristic of coronal holes, in contrast to all other regions on the Sun, where the magnetic field distribution is typically more complex, mixed, and balanced in polarity. 

The structure of this paper is as follows: Section 2 describes the datasets used in our study. Section 3 outlines the methodology and the python code employed to analyze the data. Section 4 presents the results in detail, and Section 5 completes the paper with a conclusion.

\begin{figure*}[ht]
    \centering
    \includegraphics[width=\textwidth]{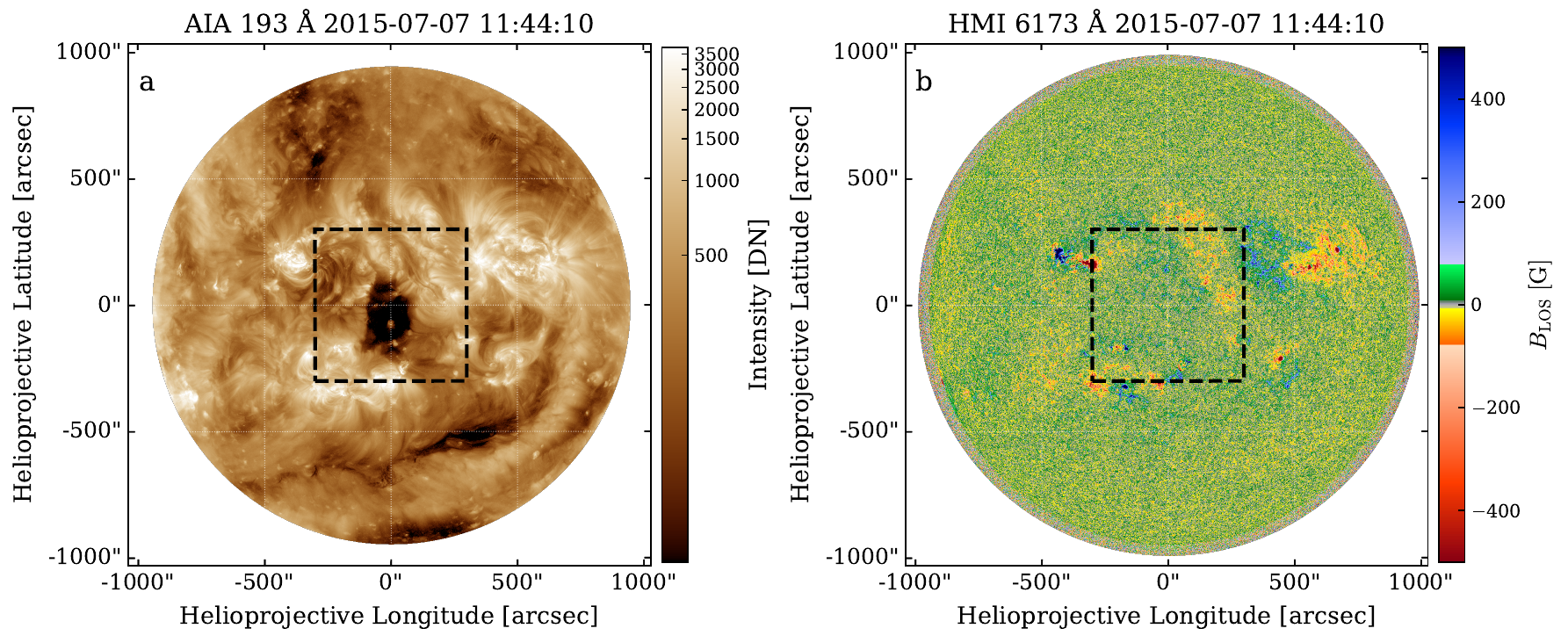}
    \caption{The left panel \texttt{`a'} displays a coronal hole observed by SDO/AIA 193~\AA{} on 2015-07-07 at 11:44:41, centered on the solar disk. The corresponding photospheric magnetogram is shown in the right panel \texttt{`b'}, as observed by SDO/HMI, depicting the line-of-sight magnetic field ($B_{\mathrm{LOS}}$). The black dotted box indicates the region of interest.}
    \label{fig:Figure_1_AIA_HMI_Aligned_ROI_map}
\end{figure*}

\section{Data} \label{sec:data}
In this study, we use both remote-sensing and in-situ datasets spanning the years 2010 to 2020, covering solar cycle 24. The remote-sensing observations are obtained from the Solar Dynamics Observatory \citep[SDO;][]{2012SoPh..275....3P}, which carries three instruments: the Helioseismic and Magnetic Imager (HMI), the Atmospheric Imaging Assembly (AIA), and the Extreme Ultraviolet Variability Explorer (EVE). SDO/AIA provides continuous observations of the solar atmosphere in $10$ different wavelength channels. We co-align data from the SDO/AIA and SDO/HMI instruments. Science-ready SDO/AIA Level 1.5 and SDO/HMI Level 1.5 data were obtained via the Virtual Solar Observatory (\href{https://nso.edu/data/vso/}{Virtual Solar Observatory, NSO})) using the \texttt{SunPy} library \citep{sunpy_community2020}. 

The corresponding in-situ solar wind plasma and magnetic field data at 1~AU were retrieved from the OMNI database (\href{https://omniweb.gsfc.nasa.gov/}{OMNIWeb database, NASA GSFC}) using the \texttt{PySPEDAS} package \citep{pyspedas_doc}. Additional details regarding data access and processing are provided in the supplementary materials \citep{katuwal2025}, which are available on Zenodo.\footnote{\url{https://doi.org/10.5281/zenodo.16521356}}
\subsection{Atmospheric imaging assembly (AIA)}
The SDO/AIA instrument captures full-disk images of the Sun in multiple bandpasses at a cadence of 12 seconds. Each full-disk image has a plate scale of $0.6\arcsec$ per pixel, with an effective spatial resolution of approximately $1.5\arcsec$ \citep{2012SoPh..275...17L}. Coronal holes are identified in the solar corona using the 193~\AA\ filtergram, which provides high contrast between the coronal holes and the surrounding non-coronal regions \citep{article,2019A&A...629A..22H}.
This filtergram shows plasma at a temperature of about 1.6 million degrees Kelvin in the Sun's corona by observing mostly Fe\,\textsc{xii} ions \citep{2021ApJ...913...28R}.
\subsection{Helioseismic and magnetic imager (HMI)}
The SDO/HMI instrument provides full-disk images of the line-of-sight photospheric magnetic flux with a with plate scale of 0.505\arcsec per pixel approximately every 45 seconds \citep{2012SoPh..275..207S}. It is derived from spectropolarimetric measurements of the Fe~\textsc{I} 6173~\AA\ spectral line. Near disk center, the photon noise level is considered to be 7 Gauss(G) \citep{2012SoPh..275..229S,2016SoPh..291.1887C}. The inverted magnetic field data contains information on the strength, polarity, and orientation of magnetic field on the surface of Sun. In addition to the line-of-sight measurements, HMI also provides a full-disk vector magnetogram derived from stokes inversions of the Fe~\textsc{I} 6173~\AA\  spectral line at a cadence of about 12 minutes \citep{2014SoPh..289.3483H}, which are not used in this study.

\subsection{In-situ data}
We used in-situ solar wind plasma and magnetic field measurements from the OMNI database (\href{https://omniweb.gsfc.nasa.gov/}{OMNIWeb database, NASA GSFC}). In this study, we used 5-minute resolution data that are time-shifted to the Earth's bow shock nose (approximately 1 AU from the Sun), ensuring uniform timing across all parameters. The plasma data include the three velocity components ($V_x$, $V_y$, $V_z$) in geocentric solar ecliptic (GSE) coordinates, from which we calculate the plasma flow speed ($V$), proton density ($N_p$), proton temperature ($T_p$), dynamic pressure ($P$), and the SYM-H geomagnetic index. The interplanetary magnetic field (IMF) data consist of the three magnetic field components ($B_x$, $B_y$, $B_z$) and the total magnetic field strength, all in GSE coordinates. In order to study high-speed streams originating from each coronal hole, we exclude time periods that show signatures of interplanetary coronal mass ejections (ICMEs) and corotating interaction regions (CIRs) by referring to \href{https://helioforecast.space}{HELIO4CAST ICME \citep{mostl2020prediction} and SIR catalog} \citep{2019JGRA..124.3871G} based on the observations from L1 monitors (Wind spacecraft). We note that the HELIO4CAST only contains the list of SIRs during the period of January 2007 -- July 2018. For events outside this time range, we applied the CIR identification criteria outlined in \citet{2006SoPh..239..337J, 2009SoPh..259..345J, 2018LRSP...15....1R, silwal2024multispecies} to identify and filter relevant intervals.

\section{Methodology} \label{sec:method}
We begin by manually examining full-disk images of the Sun 
(\href{https://suntoday.lmsal.com/suntoday}{SunToday, LMSAL}) from 2010 to 2020.
From visual inspection, we consider potential coronal hole candidates based on the following criteria: it is centered on the solar disk, within a defined area of y-axis ($-500$\arcsec, $500$\arcsec) and x-axis ($-500$\arcsec, $500$\arcsec); it must be isolated. These criteria reduce geometric effects and allow us to connect the coronal hole to high speed solar wind streams in-situ at 1 AU.
\begin{figure*}[ht]
    \centering
    \begin{subfigure}
        \centering
        \includegraphics[width=\textwidth]{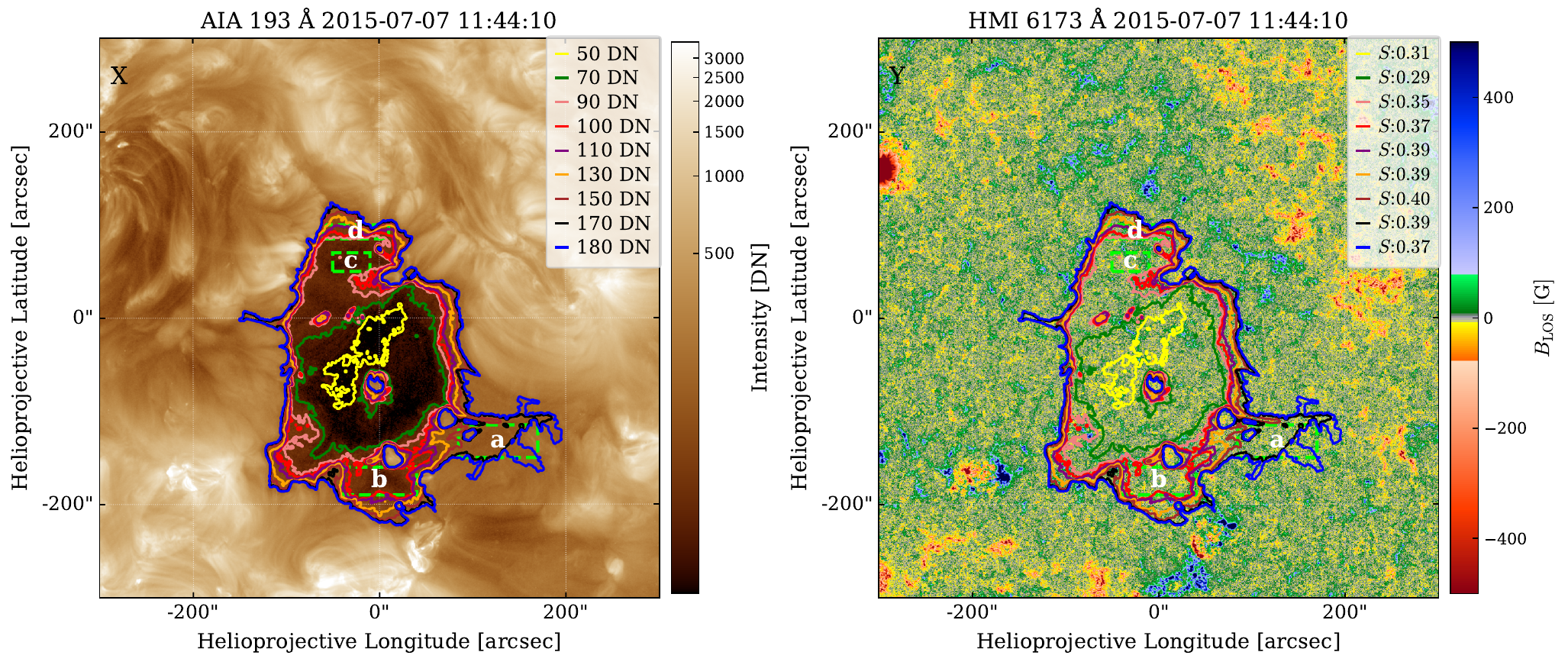}
        \caption{Left (\texttt{X}): Coronal hole boundaries defined using different intensity thresholds in the solar corona (cropped image of panel~\texttt{`a'} in Figure~\ref{fig:Figure_1_AIA_HMI_Aligned_ROI_map}). 
Right (\texttt{Y}): Overlays of all corresponding contours on the photospheric magnetogram (cropped image of panel~\texttt{`b'} in Figure~\ref{fig:Figure_1_AIA_HMI_Aligned_ROI_map}). 
The label \texttt{`a'} denotes the region of interest outside the boundary (100~DN), while \texttt{`b'}, \texttt{`c'}, and \texttt{`d'}
represent regions inside the boundary.
}     
        \label{fig:aia_hmi_varying_boundary}
    \end{subfigure}
    \begin{subfigure}
        \centering
        \includegraphics[width=\textwidth]{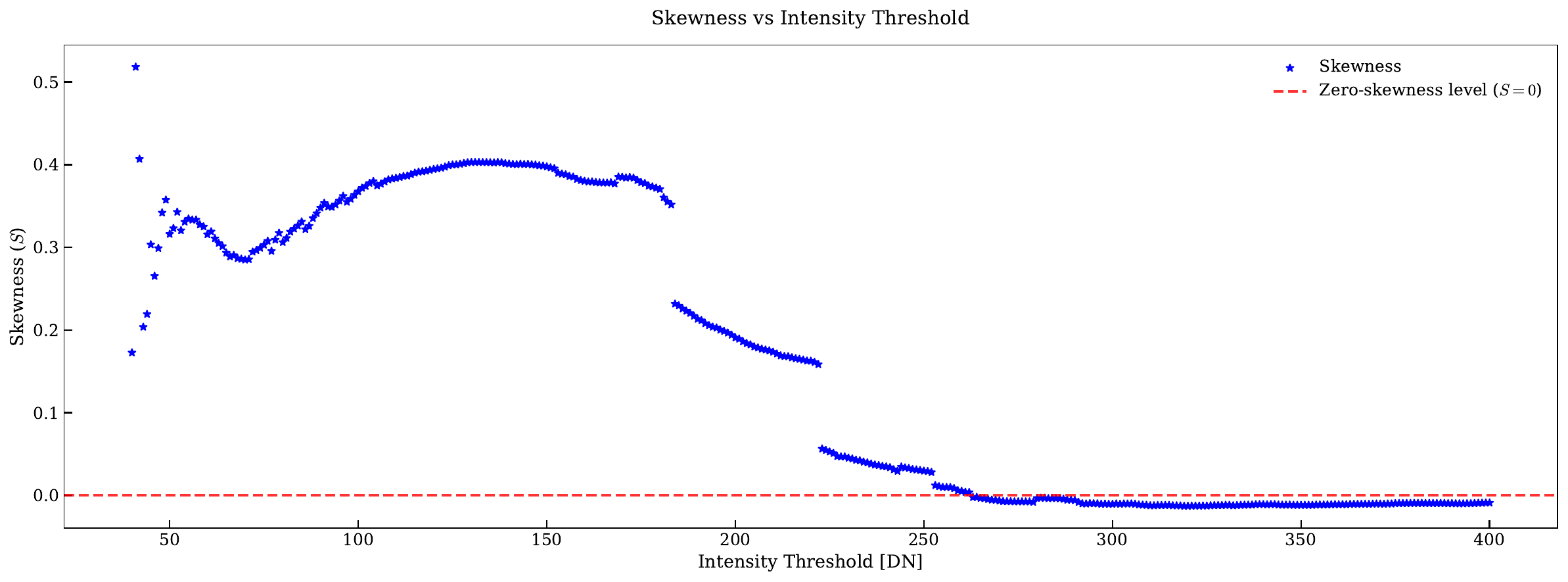}
        \caption{Skewness of magnetic unipolarity as a function of intensity threshold. The y-axis represents the skewness of $B_{\text{LOS}}$ distribution, and the x-axis represents the different intensity threshold (DN) used to define the boundaries of coronal hole. The horizontal red dashed line at $S = 0$ marks the zero-skewness level, highlighting sign changes in the skewness.
}
        \label{fig:Skewness_variation}
    \end{subfigure}
\end{figure*}

We employed the SDO/AIA 193~\AA\ filtergram to define the coronal hole boundaries in the solar corona, as shown in Figure~\ref{fig:Figure_1_AIA_HMI_Aligned_ROI_map}a. The SDO/HMI magnetogram was used to analyze $B_{\text{LOS}}$ beneath the coronal hole, at the same disk location and observation time on the photosphere, as shown in Figure~\ref{fig:Figure_1_AIA_HMI_Aligned_ROI_map}b. An algorithm developed by the SunPy community \citep{sunpy_community2020} was implemented for data acquisition and preparation, as well as for image processing and analysis. We cropped the region of interest within the same field of view of the AIA 193~\AA\ and HMI 6173~\AA\ images using the \texttt{submap()} function of SunPy, as indicated by the dotted black line in both panels of Figure~\ref{fig:Figure_1_AIA_HMI_Aligned_ROI_map}. The AIA 193~\AA\ image is aligned to the same grid as the HMI 6173~\AA\ image using the \texttt{reproject\_to()} function. We avoid rescaling the HMI magnetogram to the plate scale of the AIA 193~\AA\ image in order to prevent potential distortions caused by interpolation. It is not possible to identify a coronal hole solely based on a magnetogram of the photosphere \citep{1979SSRv...23..139H}.
Instead, coronal hole boundaries are typically defined using a variety of automated detection techniques \citep{2012SoPh..281..793R,2009SoPh..256...87K,2014A&A...561A..29V,2018JSWSC...8A...2G,2016ApJ...823...53C,2019SoPh..294..144H} applied to multiple EUV wavelengths, including 193~\AA\ , 211~\AA\, and 171~\AA\ \citep{2024ApJS..271....6R}. Among these, the AIA 193~\AA\ channel is commonly used for coronal hole detection.
\begin{figure*}[ht]
    \centering
    \begin{subfigure}
        \centering
        \includegraphics[width=\textwidth]{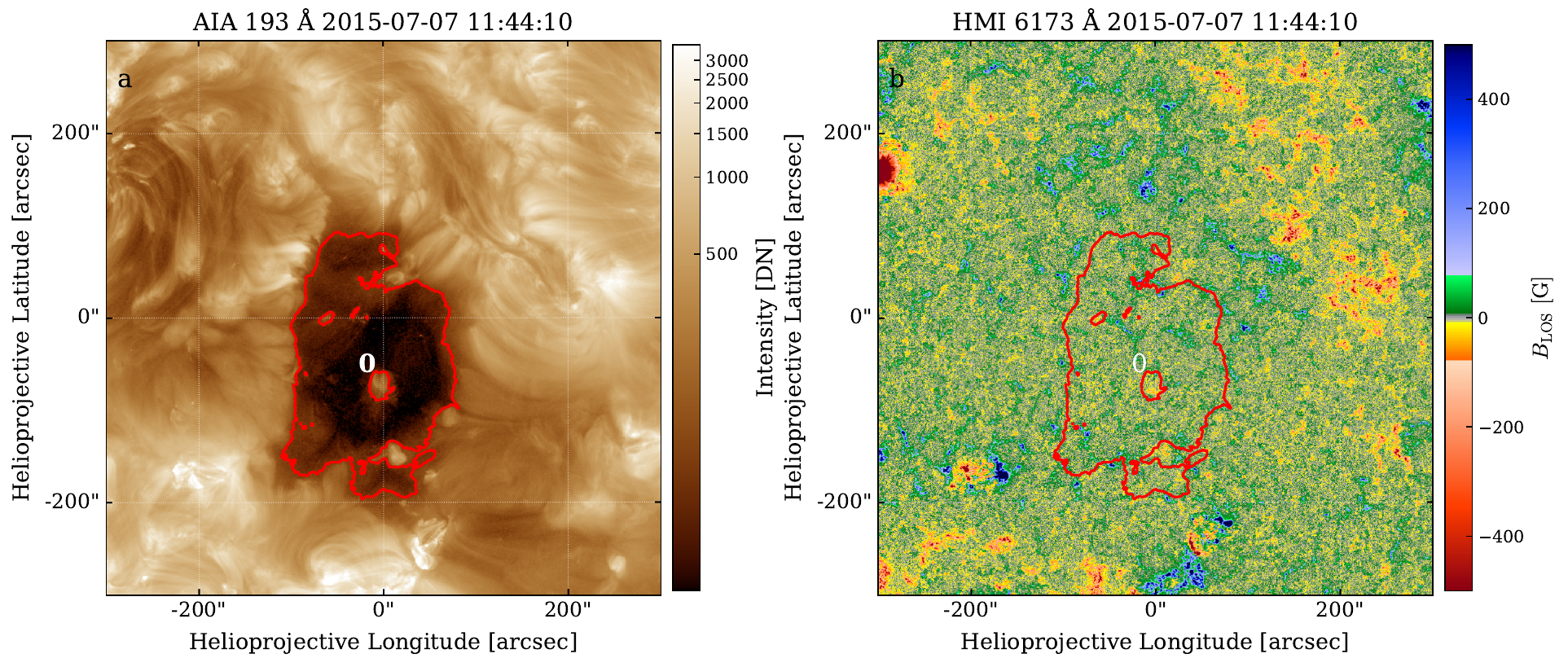}
        \label{fig:Final_coronal_hole}
    \end{subfigure}
    \begin{subfigure}
        \centering
        \includegraphics[width=\textwidth]{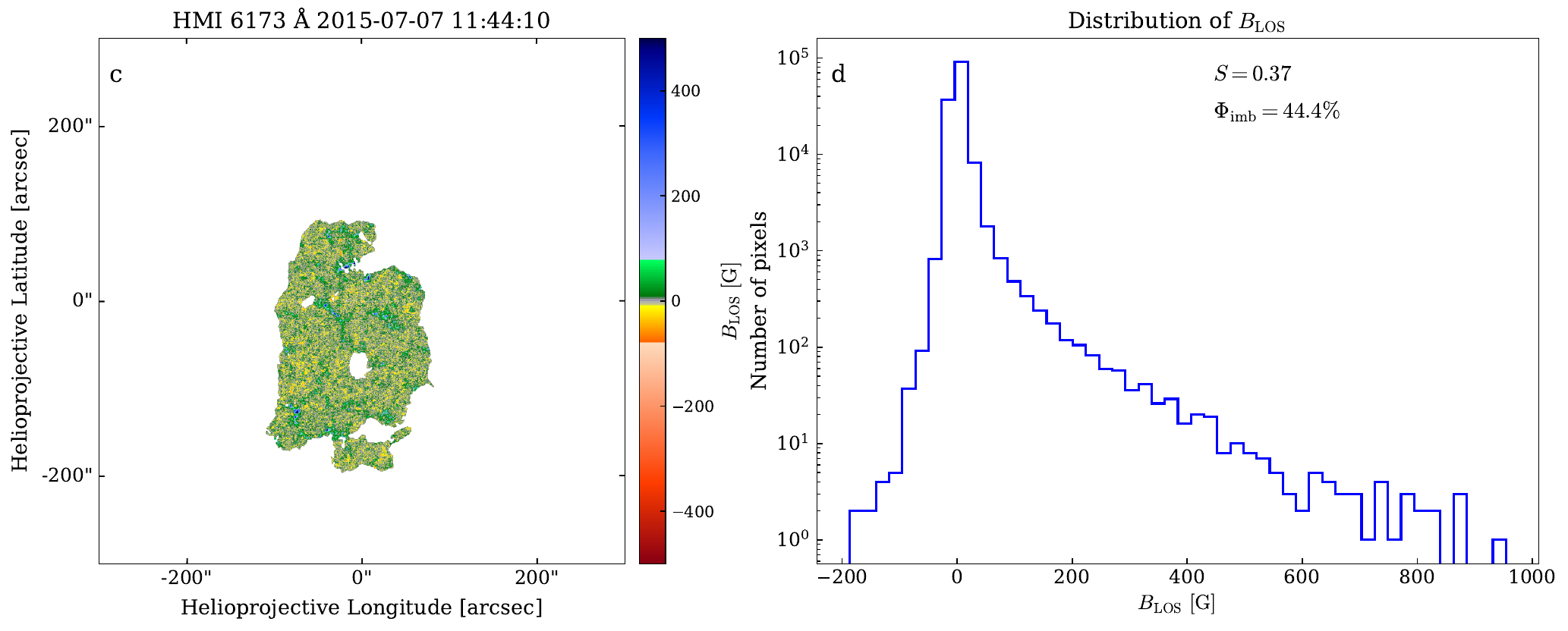}
        \caption{Top left \texttt{(`a')}: coronal hole boundary at an intensity threshold of 100~DN (`a' boundary from Figure~\ref{fig:aia_hmi_varying_boundary}X). Top right \texttt{(`b')}: The same contour overlaid on the corresponding region on the photospheric magnetogram (the corresponding boundary from Figure~\ref{fig:aia_hmi_varying_boundary}Y). In both maps (\texttt{`a'} and \texttt{`b'}), the label \texttt{0} denotes the coronal hole region enclosed by the boundary defined at 100~DN, shown in two different atmospheric layers. Bottom left \texttt{(`c')}: Final structure of the coronal hole obtained by masking the HMI magnetogram based on AIA intensity. Bottom right \texttt{(`d')}: Positively skewed $B_{\mathrm{LOS}}$ distribution corresponding to the coronal hole shown on the left.
}
        \label{fig:Coronal_map_histogram}
    \end{subfigure}
\end{figure*}

To determine the coronal hole boundary, a range of intensity thresholds was applied to the AIA 193~\AA\ image shown in Figure~\ref{fig:aia_hmi_varying_boundary}X. Different colored contours represent boundaries obtained using different intensity thresholds. These contours were overlaid on the corresponding photospheric magnetogram shown in Figure~\ref{fig:aia_hmi_varying_boundary}Y. The skewness of the $B_{\mathrm{LOS}}$ distribution was calculated for all pixels enclosed within each contour. The skewness ($S$) is computed using Karl Pearson’s coefficient:
\begin{equation}
S = \frac{3(\mu - M)}{\sigma}
\tag{1}
\end{equation}
where $\mu$, $M$, and $\sigma$ are the mean, median, and standard deviation of the $B_{\mathrm{LOS}}$ distribution, respectively. A value of $S = 0$ indicates a balanced (symmetric) magnetic flux distribution, while $S > 0$ or $S < 0$ implies a dominance of positive or negative polarity, respectively, corresponding to a unipolar magnetic region. 
Skewness provides a measure of asymmetry in the $B_{\mathrm{LOS}}$ distribution within a coronal hole, where a strongly skewed distribution indicates magnetic dominance of one polarity, consistent with the expected unipolar nature of coronal holes.
\begin{figure*}[ht]
    \centering
    \includegraphics[width=\textwidth]{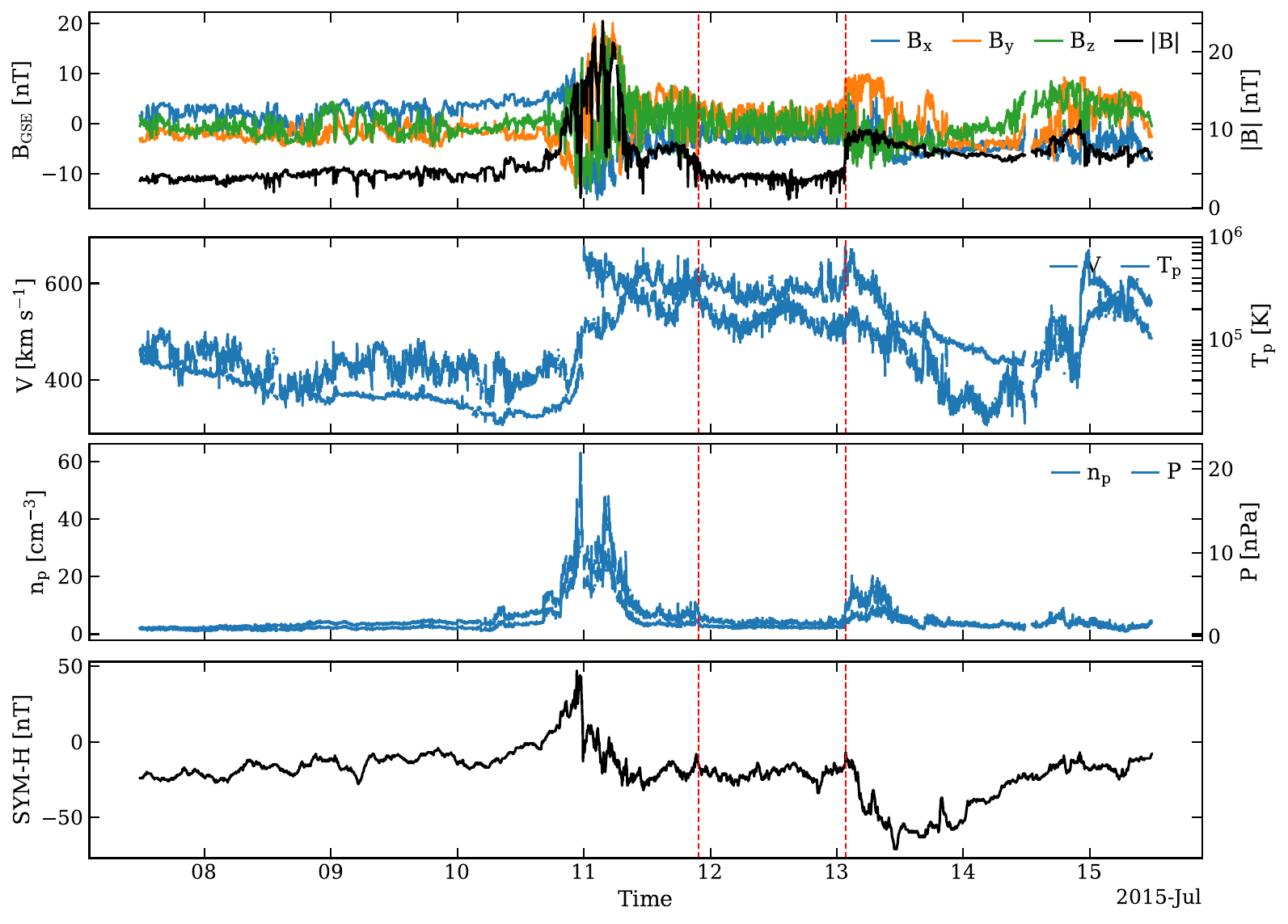}
    \caption{In-situ solar wind and magnetic field parameters associated with the coronal hole (Figure~\ref{fig:Figure_1_AIA_HMI_Aligned_ROI_map}). Panels (top to bottom) display: magnetic field components ($B_x$, $B_y$, $B_z$) with total magnitude ($|B|$), solar wind speed ($V$) with proton temperature ($T_p$), proton density ($n_p$) with pressure ($P$), and the SYM-H index. The vertical red dashed lines indicate the time interval of the constant phase of the high-speed solar wind stream used for averaging (OMNIWeb, NASA GSFC).}
    \label{fig:Insitu_Final_small}
\end{figure*}

The variation of skewness as a function of the intensity threshold used to define the coronal hole boundary is shown in Figure~\ref{fig:Skewness_variation}. The y-axis represents the skewness, and the x-axis shows the intensity threshold. A horizontal dashed line at $S = 0$ is included to indicate the zero-skewness level, allowing one to identify when and where the skewness changes sign. Based on this analysis, an intensity threshold of 100~DN was chosen to define the coronal hole boundary, as shown in Figure~\ref{fig:Coronal_map_histogram}a and overlaid on the photospheric magnetogram in Figure~\ref{fig:Coronal_map_histogram}b (for more details, see Section~4.1).  We create a mask based on the intensity (pixel DN) from the AIA 193~\AA\ image, and apply it to the HMI magnetogram to extract $B_{\text{LOS}}$ for corresponding photospheric region \citep{Mumford2020}. To isolate the region of interest (labeled \texttt{‘0’} in Figures~\ref{fig:Coronal_map_histogram}a and Figures~\ref{fig:Coronal_map_histogram}b), a bounding box was drawn around the identified contour(the red contour in Figures~\ref{fig:Coronal_map_histogram}a) and applied to the HMI magnetogram (Figures~\ref{fig:Coronal_map_histogram}b). This process effectively masks out all regions except the one labeled \texttt{`0'}, as shown Figure~\ref{fig:Coronal_map_histogram}b. The resulting magnetic map at the photospheric height below the coronal hole is shown in Figure~\ref{fig:Coronal_map_histogram}c. One example is available at \url{https://zenodo.org/records/16521356} in the file titled \textit{Maincode\_project.ipynb}.
\subsection{Magnetic flux imbalance}
After producing the final map of the photospheric magnetic field beneath the coronal hole shown in Figure~\ref{fig:Coronal_map_histogram}c, we constructed a histogram to examine the distribution of $B_{\text{LOS}}$ within the coronal hole, as shown in Figure~\ref{fig:Coronal_map_histogram}d. The presence of a skewed distribution motivated us to quantify the asymmetry using skewness and magnetic flux imbalance. The magnetic flux imbalance ($\Phi_{\mathrm{{imb}}}$) is calculated by summing all positive magnetic field values (\(B_i^+\)) and negative magnetic field values (\(B_i^-\)) separately, computing their difference, and dividing it by the total unsigned magnetic flux, \(\left| \sum B_i^+ \right| + \left| \sum B_i^- \right|\). The result is expressed as a percentage.
\begin{equation*}
\Phi_{\mathrm{imb}} =
\frac{\sum\limits_{i=1}^{N} B_i^{+} - \sum\limits_{i=1}^{N} B_i^{-}}
{\left| \sum\limits_{i=1}^{N} B_i^{+} \right| + \left| \sum\limits_{i=1}^{N} B_i^{-} \right|}
\times 100\%
\tag{2}
\end{equation*}
This formulation follows the approach of \citet{2004SoPh..225..227W,2009SoPh..260...43A,2017ApJ...835..268H}. The magnetic flux imbalance quantifies the relative contribution of the dominant magnetic polarity to the total unsigned flux within the coronal hole. Both metrics are important to study coronal holes magnetic fields because they play vital roles in the formation of high-speed solar wind.
\subsection{High-speed solar wind stream (HSS) identification}
To interpret the magnetic flux imbalance in terms of it's impact on in-situ measurements, we examine the associated HSS speed ($v_{\mathrm{HSS}}$) at 1 AU using solar wind plasma and magnetic field data from L1 monitors. Contributions from interplanetary coronal mass ejections (ICMEs) or stream interaction regions (SIRs) can contaminate HSS signatures originating from coronal holes. To eliminate such contaminations, we exclude all the HSS intervals that overlap with ICME and SIR events, based on the HELIO4CAST ICME catalog \citep{mostl2020prediction} and the SIR catalog \citep{2019JGRA..124.3871G}. We also discard ambiguous intervals containing discontinuity-like features in the solar wind parameters and IMF, which may be locally generated within the solar wind. For the remaining intervals corresponding to the events listed in our catalog, we carefully investigate the time series of solar wind parameters and magnetic field components to ensure that no residual signatures of large-scale structures (e.g., ICMEs or CIRs) remain. For instance, Figure~\ref{fig:Insitu_Final_small} shows, from top to bottom, the magnetic field components $B_x$ (nT), $B_y$ (nT), and $B_z$ (nT) in GSE coordinates, the total magnetic field magnitude $B$ (nT), the proton bulk speed $V_{sw}$ (km~s$^{-1}$), proton temperature $T_p$ (K), proton number density $N_p$ (cm$^{-3}$), total plasma pressure $P$ (nPa), and the geomagnetic index SYM-H (nT). This event corresponds to a representative clean interval of a high-speed solar wind stream (HSS) associated with the coronal hole shown in Figure~\ref{fig:Figure_1_AIA_HMI_Aligned_ROI_map}. The observed increase in solar wind speed, decrease in proton density, increase in proton temperature, and smooth magnetic field components are characteristic of a typical HSS associated with a coronal hole, with no indication of a geomagnetic activity in the SYM-H index. We then calculate the average solar wind speed during the approximately constant phase of the HSS, which is identified by visual inspection (e.g., the interval bounded by the two red dashed lines in Figure~\ref{fig:Insitu_Final_small}). The average value over this interval, hereafter referred to as $v_{\mathrm{HSS}}$, is used as the characteristic solar wind speed for each HSS event.

\section{Results} \label{sec:Result}
\subsection{Boundary of Coronal hole}
The mapping of coronal hole boundaries has been a significant focus within the solar physics community for past few decades. Various automated detection techniques \citep{2012SoPh..281..793R,2009SoPh..256...87K,2014A&A...561A..29V,2018JSWSC...8A...2G,2016ApJ...823...53C,2019SoPh..294..144H} have been developed to identify the boundaries of coronal holes using different approaches. However, the exact boundary of a coronal hole will always remain debatable. The comparison of these methods reveals substantial uncertainties in the derived properties such as area and open magnetic flux \textbf{\citep{2021ApJ...918...21L,2021ApJ...913...28R,2024ApJS..271....6R}}.

In panel \texttt{`X'} of Figure~\ref{fig:aia_hmi_varying_boundary}, each colored contour represents a distinct intensity threshold (DN) level, illustrating how the identified coronal hole boundaries change based on the chosen intensity criterion. Each of these contours are projected onto the HMI magnetogram, as shown in panel \texttt{`Y'} of Figure~\ref{fig:aia_hmi_varying_boundary}, to calculate the skewness of the \(B_{\mathrm{LOS}}\) distribution. 
A non-zero skewness (asymmetric distribution) of magnetic flux strongly suggests the presence of a unipolar magnetic field beneath the coronal hole in the photosphere. The nature of the unipolar magnetic field indicates dominance of positive magnetic flux or negative magnetic flux, and is strongly dependent on the phase of the solar cycle \citep{2016SoPh..291.2329B,2017SoPh..292...18L}. The degree of unipolarity is quantified by the magnitude of the skewness of the $B_{\mathrm{LOS}}$ distribution and the flux imbalance.
We calculate $S$ at different intensity thresholds. The skewness values at intensity thresholds (50, 70, 90, 100, 110, 130, 150, 170, and 180) DN correspond to (0.31, 0.29, 0.35, 0.37, 0.39, 0.40, 0.39, and 0.37) respectively. Not all contours drawn from 40 DN to 400 DN are displayed in both panels of Figure~\ref{fig:aia_hmi_varying_boundary}. When the boundary thresholds differ by (10–20) DN on average, the unipolarity does not change significantly, as illustrated in Figure~\ref{fig:Skewness_variation} (for the same coronal hole used in Figure~\ref{fig:aia_hmi_varying_boundary}), where the intensity threshold step size is 1 DN. We investigate how the magnetic skewness of a coronal hole depends on the chosen boundary location by systematically varying the intensity threshold used to define coronal hole boundaries from 40 DN to 400 DN, as shown in Figure~\ref{fig:Skewness_variation}. The x-axis represents the intensity threshold (DN), and the y-axis represents skewness. All pixels below the intensity threshold of 50 DN are inadequate to define the boundary of coronal holes, leading to fluctuating skewness, as seen in Figure~\ref{fig:Skewness_variation}.
This occurs because such low thresholds select only small and uneven parts of the coronal hole that do not represent its overall structure. 
To ensure a more reliable and stable analysis, we start from 50 DN and extend the intensity threshold range up to 400 DN. As the intensity threshold increases from 70 DN, the unipolarity shows a slight change; after 90 DN, it remains ordered and consistent up to 180 DN. Here, \textit{ordered} describes the weak variation in the skewness values, indicating that the degree of unipolarity remains constant across thresholds. \textit{Consistent} refers to the preservation of the same dominant polarity, with no polarity reversal within this range \citep{2017ApJ...835..268H}. This is despite the changes in shape and size of the boundary, and suggests the real coronal hole boundary is somewhere in this region, i.e, as the contour grows to encompass more area, the growth continues to include areas of dominant unipolar components. 

In this work, we define the boundary of all coronal holes at intensity threshold of 100 DN as shown Figure~\ref{fig:Coronal_map_histogram}a and Figure~\ref{fig:Coronal_map_histogram}b , but in reality the skewness value obtained is not strongly dependent on this choice. Beyond $180$ DN, the unipolarity drops significantly because the boundary begins to include non–coronal hole regions. This affects the overall magnetic flux distribution and makes it more bipolar. The presence of a small bright structure in the observation further supports the contribution from a bipolar field, highlighted by the enclosed green box (\texttt{`a'})  in both panels of Figure~\ref{fig:aia_hmi_varying_boundary}. When the intensity threshold exceeds 250~DN, the polarity even flips. Therefore, thresholds greater than 180~DN also do not accurately represent the true coronal hole boundary, as the unipolarity changes significantly.

In order to more accurately determine the exact boundary of coronal holes in future studies, it will be useful to derive additional physical properties such as plasma velocity and coronal magnetic field at the boundary. Coronal holes are characterized by fast plasma outflows along open, unipolar magnetic field lines. At their boundaries, plasma speed typically decrease \citep{2005JGRA..110.4104S,2022ApJ...931...96A}, and the magnetic fields transition from predominantly unipolar inside the hole to mixed-polarity fields outside. Detecting these gradients in velocity and magnetic topology would enable a more accurate determination of the open–closed field boundary. The difference in magnetic field configuration inside and outside the boundary suggests it is complex. One candidate for localized brightening is magnetic reconnection taking place at the point of transition to a bi-polar distribution. 
At an intensity threshold of 170 DN and 180 DN, both contours included non-coronal hole regions, marked inside the green boxes as shown in both panels of Figure~\ref{fig:aia_hmi_varying_boundary} and labeled \texttt{`a'}, \texttt{`b'}, \texttt{`c'}, and \texttt{`d'}. These represent regions of interest for studying unipolarity on a smaller scale. The regions of interest \texttt{`b'}, \texttt{`c'}, and \texttt{`d'} are located inside the 100 DN contour. Including these regions in the contour does not alter the skewness, which suggests all of these regions contribute unipolar components, whereas region \texttt{`a'}, which is outside the fixed boundary (above 100 DN), contributes a bipolar field. The distribution of the magnetic field inside these smaller-scale regions inside the coronal holes also show unipolarity individually. We find the skewness of regions \texttt{`b'}, \texttt{`c'}, and \texttt{`d'} to be 0.21, 0.21, and 0.29, respectively. This implies that the magnetic field distributions of coronal hole regions are unipolar at HMI resolution. However, whether the distribution of unipolarity inside the coronal holes is truly uniform or non-uniform remains unknown. To resolve this, we need high-resolution data that could reveal the actual distribution of the magnetic flux at their fundamental sizes. By comparison, the skewness of region \texttt{`a'} is -0.05, indicating non-unipolarity, which is to be expected of regions that lie outside the boundary of the coronal hole.
\begin{figure*}[ht]
    \centering
    \begin{subfigure}
        \centering
    \includegraphics[width=\textwidth]{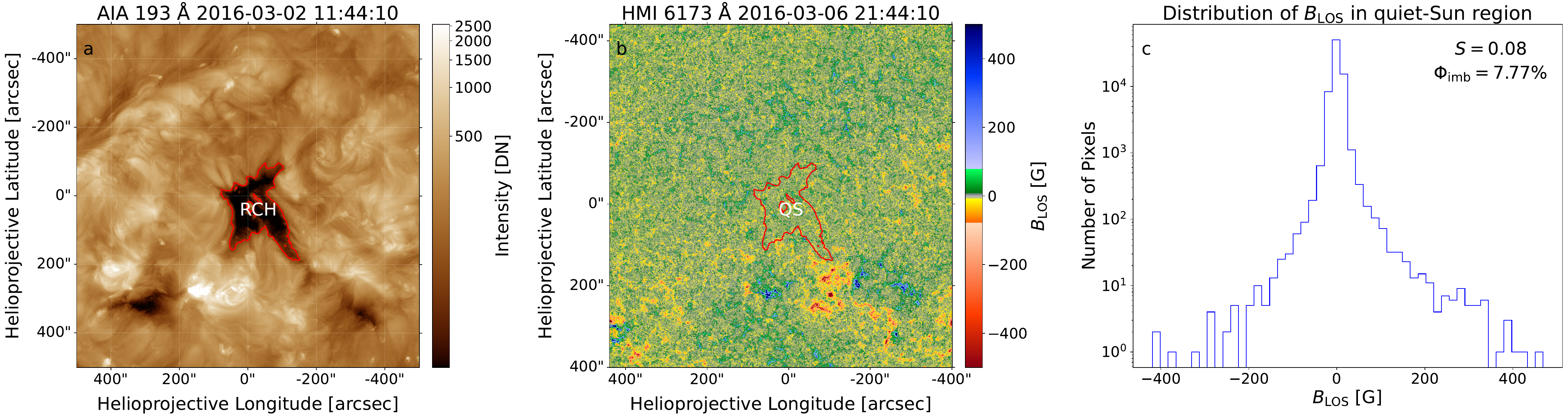}
        \caption{{Panel \texttt{`a'} marks the reference coronal hole (RCH), panel \texttt{`b'} presents an HMI magnetogram 
highlighting a quiet-Sun region with the overlaid red contour of the RCH, and panel \texttt{`c'} 
shows the corresponding symmetric distribution of $B_{\mathrm{LOS}}$.}}
        \label{fig:Quiet_SUN_combined}
    \end{subfigure}
    \begin{subfigure}
        \centering
    \includegraphics[width=\textwidth]{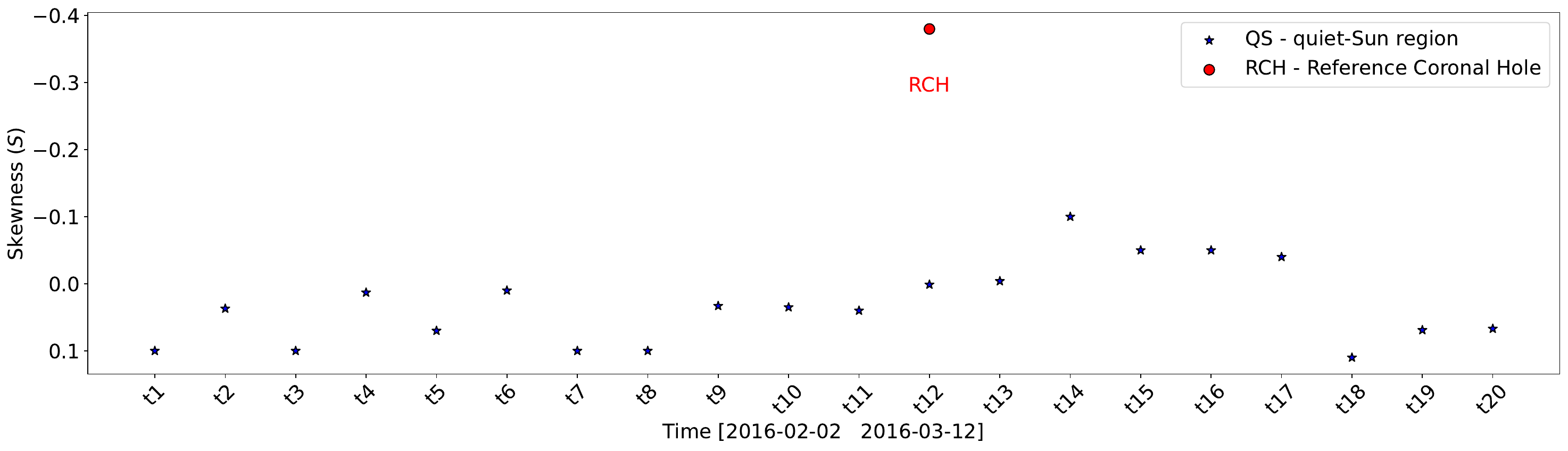} 
        \caption{The red circle (RCH) represents the reference coronal hole (same as in Figure ~\ref{fig:Quiet_SUN_combined}a, while the blue star (*) marks the quiet-Sun regions. This plot compares the magnetic field distribution in quiet-Sun regions of equal size to that of the coronal hole (RCH).}
        \label{fig:QS_RCH_Skewness}
    \end{subfigure}
\end{figure*}
\begin{figure*}[ht]
    \includegraphics[width=1.0\textwidth]{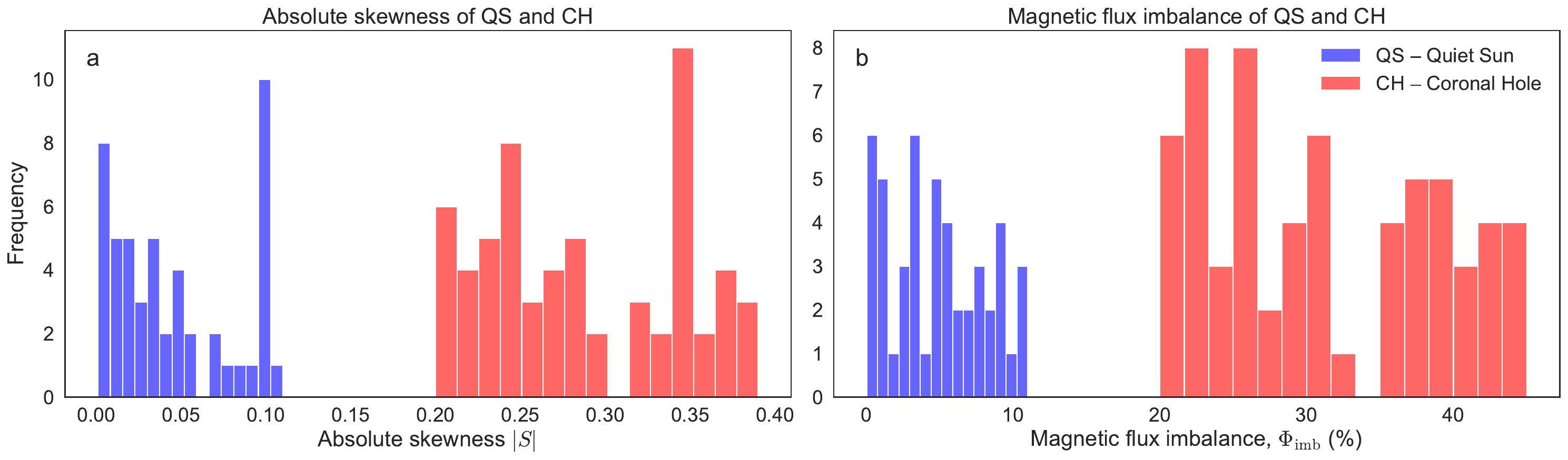}
    \caption{Left: Histogram of skewness distribution for the quiet sun regions (blue) and coronal holes (red). This graph illustrates how skewness varies between these two solar features. Right: Histogram of magnetic flux imbalance for the quiet sun regions(blue) and coronal holes (red). This graph shows the differences in magnetic flux imbalance between these regions.}    
    \label{fig:imbalance_flux_range_skew_QS_CH}
\end{figure*}
\subsection{Flux imbalance of coronal holes and quiet-Sun regions}
Coronal holes can be distinguished from the quiet-Sun regions in EUV observations of the solar corona because they emit lower radiation at coronal temperatures \citep{2004SoPh..225..227W}. However, distinguishing them in the lower solar atmosphere, such as the photosphere and chromosphere, is more challenging due to the lower contrast between densities at these cooler temperatures. Although sunspots are visually discernible in HMI magnetogram and can be easily excluded from our region of interest in the photosphere, the quiet sun  regions and coronal holes appear similar in the HMI magnetogram at the given resolution. However, understanding the magnetic field distribution is essential for differentiating coronal holes from the quiet-Sun regions.

The non-Gaussian distribution of the magnetic field within a coronal hole is a key indicator of its predominantly unipolar nature. Figure~\ref{fig:Coronal_map_histogram}c shows the $B_{\mathrm{LOS}}$ map of the coronal hole, where mixed polarities are visually present but do not clearly reveal polarity dominance. To quantify this, we plotted the histogram in Figure~\ref{fig:Coronal_map_histogram}d, which depicts the spread and asymmetry of $B_{\mathrm{LOS}}$ values. We find a skewness of $0.37$ (positive polarity) and a flux imbalance of $44.4\%$, confirming that one polarity strongly dominates within the coronal hole. This magnetic flux imbalance is consistent with the findings of \citet{2017ApJ...835..268H}.
The histogram is the basis for calculating the skewness, which shows polarity dominance.
 Different methods have been used to test polarity by calculating skewness. We compare our coronal hole candidates with previous studies \citep{Reiss_2021,2014ApJ...783..142L} and find that they have the same polarity. In addition, we study the coronal hole candidates listed in \citep{2023RLSFN..34.1045C}, where 14 out of 60 candidates are similar to ours candidates and also showed the same polarity. These results on the polarity are consistent with our findings. A similar analysis was performed to study the magnetic skewness and magnetic flux imbalance from $70$ disk-centered coronal holes. We find that $88\%$ of these coronal holes exhibit skewness values ranging from $\pm (0.20~\text{to}~0.40)$, and the magnetic flux imbalance varies between $(20\%~\text{to}~45\%)$. This result is consistent with the findings reported in the literature \citep{2009SoPh..260...43A,2017ApJ...835..268H}. These variations in skewness and magnetic flux imbalance arise due to differences in the formation timing, location, size, and magnetic field strength of coronal holes observed throughout solar cycle 24. The typical ranges of skewness  and magnetic flux imbalance offer a valuable diagnostic for characterizing the unipolarity of coronal holes.

In Figure~\ref{fig:Quiet_SUN_combined}a, RCH refers to the reference coronal hole located near the center of the solar disc, observed on 2016-03-02 11:44:41. The skewness for this region is found to be $-0.38$ (negative polarity dominates). Refer to candidate~54 in the supplementary file \textit{Coronal\_Candidates\_70.pdf} (\url{https://zenodo.org/records/16521356}).\ We project this coronal hole (RCH) onto the quiet-Sun region shown in Figure~\ref{fig:Quiet_SUN_combined}b, at the same spatial location but observed at a later time (2016-03-06 21:44:41). Quiet-Sun regions typically exhibit a bipolar distribution of $B_{\mathrm{LOS}}$
\citep{2010ApJ...719..131I,2012A&A...548A..62H}, where positive and negative polarities are nearly balanced, resulting in a symmetric histogram as shown in Figure~\ref{fig:Quiet_SUN_combined}c; we find a skewness of $0.08$ and a magnetic flux imbalance of $7.7\%$. Our analysis of the magnetic flux imbalance in quiet-Sun regions agrees with \citep{2004SoPh..225..227W}, who reported an average imbalance of $\approx 9\%$. We analyzed additional quiet-Sun regions to strengthen the robustness and accuracy of our results. 
In Figure~\ref{fig:QS_RCH_Skewness}, the RCH denotes the reference coronal hole (same as in Figure~\ref{fig:Quiet_SUN_combined}a).  The red contour corresponding to the RCH is overlaid on quiet-Sun regions of similar size, observed at earlier times $(t_1, t_2, \dots, t_{11})$ and later times $(t_{12}, t_{13}, \dots, t_{20})$, but at the same location. The x-axis marks the observation times of the quiet-Sun regions, which are not uniformly spaced. The y-axis shows the corresponding skewness values, which predominantly fall within the range of $\pm(0.05~\text{to}~0.11)$. In total, we analyzed $48$ quiet-Sun regions: $20$ from our time-series observations $(t_1$--$t_{20})$ and an additional $28$ regions from datasets provided in \citet{2023RLSFN..34.1045C}. We find that the magnetic skewness in quiet-Sun regions ranges from $\pm(0.001~\text{to}~0.11)$, and the magnetic flux imbalance varies between $(0.037\% $ to $~11.0\%)$. 

Figure~\ref{fig:imbalance_flux_range_skew_QS_CH} illustrates the distinct magnetic characteristics of quiet-Sun regions (blue) and coronal holes (red). The left panel shows that coronal holes  exhibit a broader and systematically higher magnetic skewness compared to quiet-Sun regions, indicating a stronger dominance of one magnetic polarity. The right panel demonstrates that coronal holes also possess a significantly larger magnetic flux imbalance, consistent with the presence of open magnetic field structures. Based on this statistical analysis, the results confirm that the magnetic field within coronal holes is intrinsically more asymmetric and unipolar than in the quiet-Sun, indicating that the magnetic field distribution in coronal holes differs significantly from that of quiet-Sun regions. For more details on the quiet-Sun regions and coronal hole candidates, refer to the supplementary materials available at \url{https://zenodo.org/records/16521356}.
\begin{figure*}[ht]
    \includegraphics[width=1.0\textwidth]{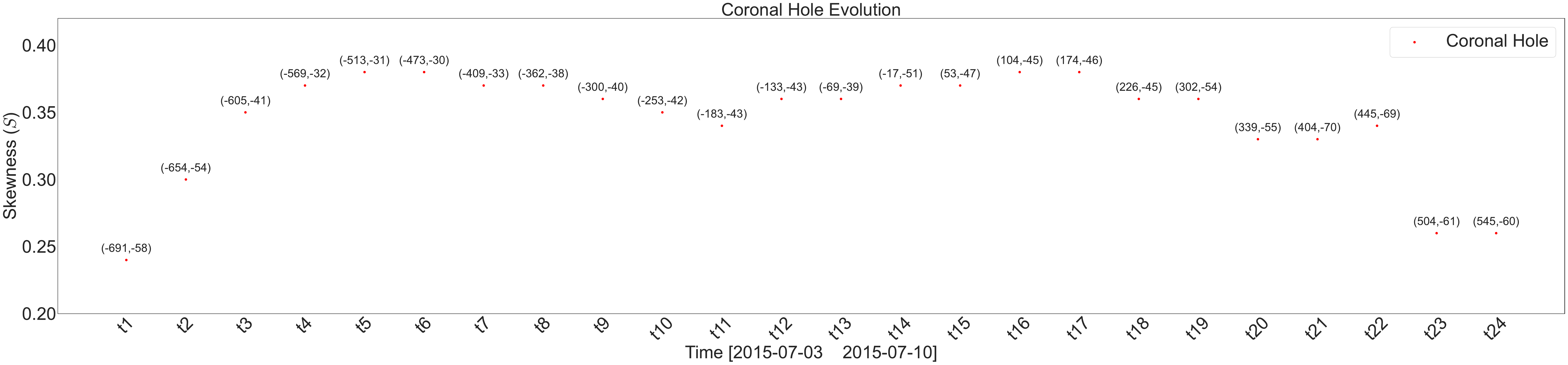}
    \caption{An asymmetric distribution of the photospheric \(B_{\mathrm{LOS}}\) within the coronal hole is observed across the solar disk, spanning from 2015-07-03 (near the east limb) to 2015-07-10 (near the west limb). Each red dot, with coordinates in arcseconds, represents the centroid position of the coronal hole as it moves across the disk.
}    
    \label{fig:CH_Evolution}
\end{figure*}
\subsection{Variation of skewness of coronal hole across disc}
We track a coronal hole from July 4, 2015, at 06:44:40 (near the east limb of the disk) to July 09, 2015, at 23:44:40 (near the west limb of the disc), as shown in Figure~\ref{fig:CH_Evolution}. The y-axis of the plot represents the skewness range, while the x-axis denotes time.\ Each data point ($t_1$, $t_2$, \dots, $t_{24}$) corresponds to an image of the same coronal hole, captured at 5-7 hour intervals over a period of approximately 6 days. To complement this analysis, we provide a supplementary movie (\textit{Evolution\_east\_to\_west\_with\_plot.mp4},
\href{https://doi.org/10.5281/zenodo.16521356}{https://doi.org/10.5281/zenodo.16521356}) that illustrates the temporal evolution of the coronal hole as it traverses the solar disk. The movie shows variations in it's shape, size, magnetic field distribution, and the dynamics of coronal bright points, including their emergence and fading, with each frame corresponding to the data points shown in Figure~\ref{fig:CH_Evolution}.\ Our aim here is to study the nature of the unipolarity of coronal holes as they traverse the disk. Variations in unipolarity provide the insight into the stability of a coronal hole's magnetic structure. The red dots with coordinates in arcsec in  Figure~\ref{fig:CH_Evolution} mark the center of the coronal hole. The coronal hole first appears at time $t_1$ on 2015-07-04 at 06:44:40 and disappears after $t_{24}$ on 2015-07-09 at 23:44:40. Before $t_1$ and after $t_{24}$, the full features of the coronal hole are not clearly visible because they are located on the far side of the Sun, out of view from our perspective. Near the disk center, we are able to fully observe the coronal hole. 

In Figure~\ref{fig:CH_Evolution}, we observe that skewness remains very close between $t_3$ and $t_{22}$, a period of approximately 5 days.  The variation before $t_3$ and after $t_{22}$ is likely due to projection effects and partial visibility, which can distort the magnetic field measurements, thereby causing the measured skewness
to drop. To reduce foreshortening effects and ensure precise and accurate magnetic field distribution measurements, in the rest of this study, we only select our coronal hole candidates from coordinates \(x = -500\arcsec\) to \(500\arcsec\) and \(y = -500\arcsec\) to \(500\arcsec\). 

There is some small variation across the central portions of the disc, including an apparent slight drop of magnetic unipolarity near the solar disk center at \(t_{12} \). Close to the disk center, these drops cannot be due to projection or foreshortening effects.\ At \(t_{12}\), the magnetic skewness is 0.36 (Positive polarity dominates), whereas before and after \(t_{12}\) skewness exhibits minor fluctuations. Analysis of the temporal evolution of the coronal hole (from the movie \textit{Evolution\_east\_to\_west\_with\_plot.mp4} is available at \url{https://zenodo.org/records/16521356}) reveals that no coronal bright points are present at $t_{12}$; however, several bright points appear and disappear within the coronal hole both before and after this time(\(t_{12}\)). These transient mixed-polarity features are the most plausible explanation for the observed slightly variations in magnetic unipolarity. The appearance and disappearance of coronal bright points locally perturb the magnetic field and can cause slight, short-lived variations in unipolarity, but they do not significantly alter the global magnetic flux distribution of the coronal hole \citep{2012A&A...548A..62H}. 
Overall, skewness and polarity of imbalance remains mostly constant \citep{2017ApJ...835..268H}, meaning coronal holes remain magnetically stable for at least half a solar rotation. In future tracking the return of a coronal hole over a few solar rotations will extend our understanding of this level of stability. Across all coronal holes studied, skewness stays within $\pm(0.20\text{ to }0.40)$, supporting our conclusion that this range is a typical characteristic of coronal holes.
\subsection{High speed stream(HSS) speed, skewness and flux imbalance}
\begin{figure*}[ht]
    \includegraphics[width=1.0\textwidth]{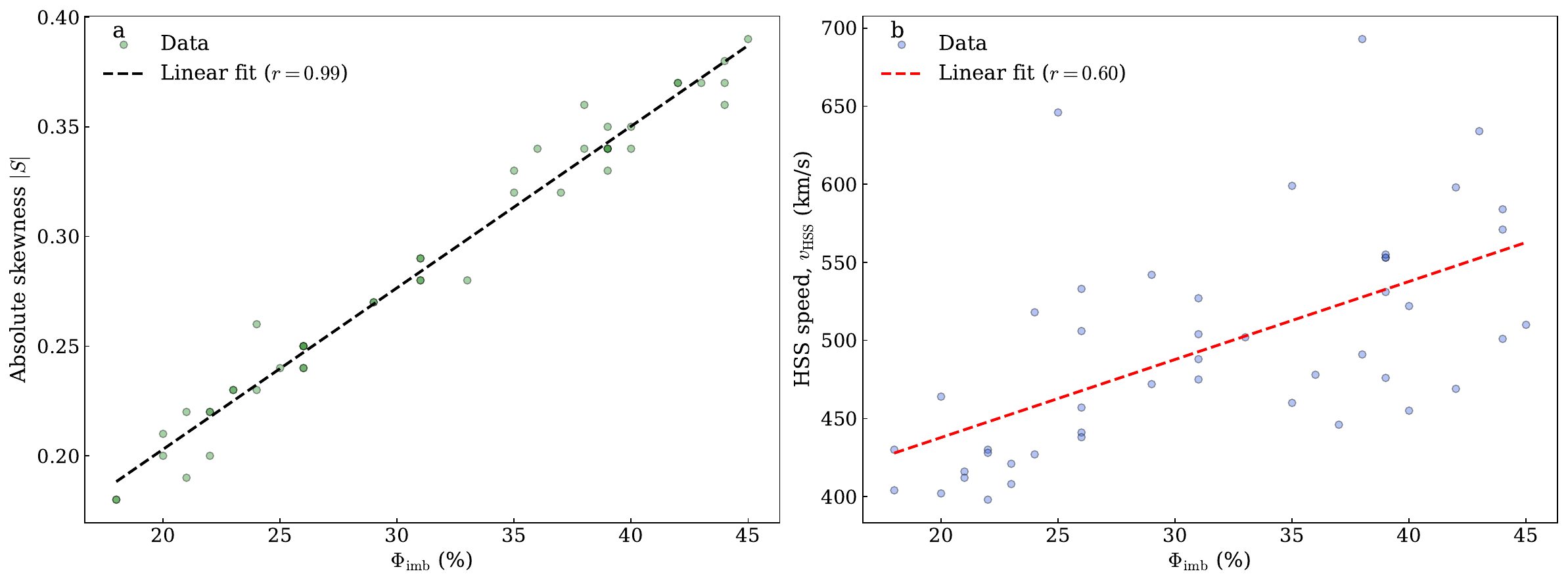}
    \caption{The left panel \texttt{`a'} shows the correlation plot between absolute magnetic skewness and 
    magnetic flux imbalance, while the right panel \texttt{`b'} presents the correlation plot between 
    magnetic flux imbalance and the HSS speed ($v_{\mathrm{HSS}}$)} measured at 1~AU.
    \label{fig:Solar_Wind}
\end{figure*}
The imbalance in magnetic flux leads to open magnetic field lines, which plays a crucial role in the formation of fast solar wind \citep{2023RLSFN..34.1045C}.
For each coronal hole in our dataset, we calculate the skewness and magnetic flux imbalance, matching these with the corresponding HSS speed ($v_{\mathrm{HSS}}$) measured at 1~AU using OMNI data (\href{https://omniweb.gsfc.nasa.gov/}{OMNIWeb database, NASA GSFC}). The \textit{in-situ} signatures at 1~AU are crucial for validating coronal hole candidates. Therefore, we analyzed various plasma parameters and magnetic field signatures measured at 1~AU.
Our goal is to correlate each coronal hole's magnetic flux imbalance with the HSS speed ($v_{\mathrm{HSS}}$). We calculated magnetic flux imbalance from the photospheric magnetogram at the time when the coronal hole is at the central meridian (black dotted box in Figure~\ref{fig:Figure_1_AIA_HMI_Aligned_ROI_map}b), yielding one value per event. The constant phase (indicated by the two vertical dashed lines in Figure~\ref{fig:Insitu_Final_small}) of the high-speed stream can occur at any time within 1.5--7~days after the coronal hole crosses the central meridian \citep{2018JGRA..123.1738H}, and this interval is used to determine HSS speed.

We extracted a subset of $46$ coronal holes to investigate the connection between magnetic flux imbalance and HSS speed. From the \textit{in-situ} measurements, some of the coronal hole signatures at 1~AU were contaminated by ICMEs and SIRs, with certain ICME signatures overlapping with coronal hole occurrences. Of the 70 coronal hole candidates, only 46 exhibited clear, uncontaminated high-speed solar wind signatures associated with coronal holes; the remaining events were affected by ICME or SIR contamination as described in Section 2.3 and 3.2. We first examined the Pearson correlation between skewness and magnetic flux imbalance for the $46$ coronal holes. The left panel 'a' of 
Figure~\ref{fig:Solar_Wind}
shows a strong Pearson correlation ($r = 0.99$) between absolute skewness, {$|S|$ (y-axis) and magnetic flux imbalance, 
$\Phi_{\mathrm{imb}}$ (x-axis).
Both parameters quantify magnetic unipolarity, and analyzing them together ensures the precision and robustness of our results. Since both describe unipolarity, we used flux imbalance ($\Phi_{\mathrm{imb}}$), which more directly traces the open magnetic flux driving the solar wind. So, we derive the relationship between the HSS speed  ($v_{\mathrm{HSS}}$) at 1~AU and magnetic flux imbalance ($\Phi_{\mathrm{imb}}$), as shown in panel 'b' of Figure~\ref{fig:Solar_Wind}. We find a moderate positive correlation ($r = 0.60$) between HSS speed  ($v_{\rm HSS}$) and magnetic flux imbalance ($\Phi_{\rm imb}$), indicating that magnetic unipolarity is one of several factors influencing high-speed stream speeds.

\section{Conclusions} \label{sec:Con}
In this study, we conducted a comprehensive statistical investigation of coronal holes, focusing on defining their boundaries, assessing the stability of their magnetic unipolarity as they traverse from one solar limb to the other, and quantifying the ranges of skewness and magnetic flux imbalance. Furthermore, we examined the relationship between magnetic flux imbalance and the HSS speed measured at 1~AU. Based on our analysis, we conclude the following:
\begin{enumerate}
   \item The unipolarity of coronal holes in terms of skewness ($S$) depends on the chosen boundary location, and this can be evaluated by systematically varying an intensity threshold as illustrated in Figure~\ref{fig:aia_hmi_varying_boundary} and Figure~\ref{fig:Skewness_variation}.
    
    \item The magnetic unipolarity within coronal holes remains constant over a few days as shown in Figure~\ref{fig:CH_Evolution}, 
    indicating that the magnetic distribution of coronal holes is persistent during their disk passage.
    
    \item The typical range of magnetic skewness in coronal holes is $\pm(0.20~\text{to}~0.40)$, 
    whereas quiet-Sun regions exhibit values of $\pm(0.001~\text{to}~0.11)$. Similarly, the 
    magnetic flux imbalance in coronal holes ranges from $20\%$ to $45\%$, compared to $0.037\%$ 
    to $11\%$ in quiet-Sun regions. These distinctions clearly differentiate the $B_{\mathrm{LOS}}$ 
    distribution in coronal holes from that in quiet-Sun regions as shown in Figure~\ref{fig:imbalance_flux_range_skew_QS_CH} .
    
    \item $S$ and $B_{\mathrm{LOS}}$ are essentially equivalent measures of 
    magnetic unipolarity as shown in Figure~\ref{fig:Solar_Wind}a. Studying either of these parameters is sufficient for quantifying 
    unipolarity; however, analyzing both provides robustness and consistency.
    
    \item We find a moderate Pearson correlation coefficient ($r = 0.60$) between the magnetic flux imbalance and the HSS speed, at 1~AU as shown in Figure~\ref{fig:Solar_Wind}b , suggesting that flux imbalance is an important factor associated with HSS speed, but its influence is partial, indicating that other physical mechanisms also shape the resulting solar wind stream.
 
\end{enumerate}

Regions dominated by unipolar magnetic fields are expected to facilitate plasma acceleration 
to higher speeds, enabling plasma to escape from the coronal holes. In contrast, regions with closed 
magnetic field lines trap plasma,
resulting in reduced outflow and slower solar wind speeds. This indicates that the magnetic field distribution plays a vital role in accelerating plasma originating from coronal holes. 
A detailed study involving solar wind modeling is required to further investigate this process.
Our findings provide an additional diagnostic for identifying coronal hole. Any dark patch with magnetic skewness in the range $\pm(0.20~\text{to}~0.40)$ and a magnetic flux imbalance
between $20\%$ and $45\%$ can be confidently classified as a coronal hole. It is important to 
note that not all dark patches observed in extreme ultraviolet and soft X-ray solar disk images represent coronal holes. Other dark features, such as coronal voids \citep{2023A&A...678A.196N} 
and filaments \citep{2014CEAB...38...95R}, also appear in the solar corona and exhibit bipolar 
magnetic fields \citep{2023A&A...678A.196N}. Filaments and coronal voids are not included in the present analysis; however, we plan to investigate them in a future study. Our analysis of coronal hole unipolarity supports 
the theoretical definition that coronal holes are regions associated with open magnetic field lines.\\

All datasets, analysis code, and results presented in this work are publicly available and can be accessed at Zenodo: \url{https://zenodo.org/records/16521356}.
\section{Acknowledgments} \label{sec:acknowledgments}

This research is supported by the DKIST Ambassadors program, funded by the National Solar Observatory through NSF Cooperative Support Agreement AST-1400450. While this work is not directly connected to DKIST, we are grateful for the program's financial assistance during the course of this study.

We obtained the data used in this study through the Virtual Solar Observatory (VSO), with additional contributions from NASA/SDO. OMNI data were accessed via the GSFC/SPDF OMNIWeb interface (\url{https://omniweb.gsfc.nasa.gov}).
The authors thank Dr. Ashok Silwal and Dr. Nishu Karna for helpful discussions on in-situ observations from L1 monitors.

\bibliography{Citefile}

@incollection{CRANMER20023,
title = {Coronal holes and the solar wind},
editor = {Petrus C.H. Martens and David P. Cauffman},
series = {COSPAR Colloquia Series},
publisher = {Pergamon},
volume = {13},
pages = {3-12},
year = {2002},
booktitle = {Multi-wavelength Observations of Coronal Structure and Dynamics},
issn = {0964-2749},
doi = {https://doi.org/10.1016/S0964-2749(02)80003-8},
url = {https://www.sciencedirect.com/science/article/pii/S0964274902800038},
author = {S.R. Cranmer}
}

@ARTICLE{2022ApJ...932..115S,
       author = {{Schonfeld}, Samuel J. and {Henney}, Carl J. and {Jones}, Shaela I. and {Arge}, Charles N.},
        title = "{Solar Polar Flux Redistribution Based on Observed Coronal Holes}",
      journal = {\apj},
         year = 2022,
        month = jun,
       volume = {932},
       number = {2},
          eid = {115},
        pages = {115},
          doi = {10.3847/1538-4357/ac6ba1},
archivePrefix = {arXiv},
       eprint = {2204.13676},
 primaryClass = {astro-ph.SR},
       adsurl = {https://ui.adsabs.harvard.edu/abs/2022ApJ...932..115S}
}

@ARTICLE{1973SoPh...29..505K,
       author = {{Krieger}, A.~S. and {Timothy}, A.~F. and {Roelof}, E.~C.},
        title = "{A Coronal Hole and Its Identification as the Source of a High Velocity Solar Wind Stream}",
      journal = {\solphys},
         year = 1973,
        month = apr,
       volume = {29},
       number = {2},
        pages = {505-525},
          doi = {10.1007/BF00150828},
       adsurl = {https://ui.adsabs.harvard.edu/abs/1973SoPh...29..505K}
}

@article{wang1990solar,
  title={Solar wind speed and coronal flux-tube expansion},
  author={Wang, Y-M and Sheeley Jr, NR},
  journal={Astrophysical Journal, Part 1 (ISSN 0004-637X), vol. 355, June 1, 1990, p. 726-732. Research supported by the US Navy.},
  volume={355},
  pages={726--732},
  year={1990}
}

@article{mccomas2007understanding,
  title={Understanding coronal heating and solar wind acceleration: Case for in situ near-Sun measurements},
  author={McComas, DJ and Velli, Marco and Lewis, WS and Acton, LW and Balat-Pichelin, M and Bothmer, Volker and Dirling Jr, RB and Feldman, WC and Gloeckler, G and Habbal, SR and others},
  journal={Reviews of Geophysics},
  volume={45},
  number={1},
  year={2007},
  publisher={Wiley Online Library}
}

@ARTICLE{1976SoPh...46..303N,
       author = {{Nolte}, J.~T. and {Krieger}, A.~S. and {Timothy}, A.~F. and {Gold}, R.~E. and {Roelof}, E.~C. and {Vaiana}, G. and {Lazarus}, A.~J. and {Sullivan}, J.~D. and {McIntosh}, P.~S.},
        title = "{Coronal holes as sources of solar wind.}",
      journal = {\solphys},
         year = 1976,
        month = feb,
       volume = {46},
       number = {2},
        pages = {303-322},
          doi = {10.1007/BF00149859},
       adsurl = {https://ui.adsabs.harvard.edu/abs/1976SoPh...46..303N}
}

@ARTICLE{2006SSRv..124...51S,
       author = {{Schwenn}, R.},
        title = "{Solar Wind Sources and Their Variations Over the Solar Cycle}",
      journal = {\ssr},
         year = 2006,
        month = jun,
       volume = {124},
       number = {1-4},
        pages = {51-76},
          doi = {10.1007/s11214-006-9099-5},
       adsurl = {https://ui.adsabs.harvard.edu/abs/2006SSRv..124...51S}
}

@ARTICLE{2004ApJ...612.1171W,
       author = {{Woo}, Richard and {Habbal}, Shadia Rifai and {Feldman}, Uri},
        title = "{Role of Closed Magnetic Fields in Solar Wind Flow}",
      journal = {\apj},
         year = 2004,
        month = sep,
       volume = {612},
       number = {2},
        pages = {1171-1174},
          doi = {10.1086/422799},
       adsurl = {https://ui.adsabs.harvard.edu/abs/2004ApJ...612.1171W}
}

@ARTICLE{1972ApJ...176..511M,
       author = {{Munro}, Richard H. and {Withbroe}, George L.},
        title = "{Properties of a Coronal ``hole'' Derived from Extreme-Ultraviolet Observations}",
      journal = {\apj},
         year = 1972,
        month = sep,
       volume = {176},
        pages = {511},
          doi = {10.1086/151653},
       adsurl = {https://ui.adsabs.harvard.edu/abs/1972ApJ...176..511M}
}

@article{altschuler1972coronal,
  title={Coronal holes},
  author={Altschuler, Martin D and Trotter, Dorothy E and Orrall, Frank Q},
  journal={Solar Physics},
  volume={26},
  pages={354--365},
  year={1972},
  publisher={Springer}
}

@ARTICLE{2014ApJ...783..142L,
       author = {{Lowder}, C. and {Qiu}, J. and {Leamon}, R. and {Liu}, Y.},
        title = "{Measurements of EUV Coronal Holes and Open Magnetic Flux}",
      journal = {\apj},
         year = 2014,
        month = mar,
       volume = {783},
       number = {2},
          eid = {142},
        pages = {142},
          doi = {10.1088/0004-637X/783/2/142},
archivePrefix = {arXiv},
       eprint = {1502.06038},
 primaryClass = {astro-ph.SR},
       adsurl = {https://ui.adsabs.harvard.edu/abs/2014ApJ...783..142L}
}

@article{cranmer2009coronal,
  title={Coronal holes},
  author={Cranmer, Steven R},
  journal={Living Reviews in Solar Physics},
  volume={6},
  number={1},
  pages={1--66},
  year={2009},
  publisher={Springer}
}

@ARTICLE{1974ApJ...194L.115H,
       author = {{Huber}, M.~C.~E. and {Foukal}, P.~V. and {Noyes}, R.~W. and {Reeves}, E.~M. and {Schmahl}, E.~J. and {Timothy}, J.~G. and {Vernazza}, J.~E. and {Withbroe}, G.~L.},
        title = "{Extreme-Ultraviolet Observations of Coronal Holes: Initial Results from SKYLAB}",
      journal = {\apjl},
         year = 1974,
        month = dec,
       volume = {194},
        pages = {L115},
          doi = {10.1086/181682},
       adsurl = {https://ui.adsabs.harvard.edu/abs/1974ApJ...194L.115H}
}

@incollection{Kucera2002,
  author    = {Therese Kucera and Carol Jo Crannell},
  title     = {Solar Physics},
  editor    = {Robert A. Meyers},
  booktitle = {Encyclopedia of Physical Science and Technology (Third Edition)},
  publisher = {Academic Press},
  year      = {2002},
  pages     = {127-147},
  isbn      = {9780122274107},
  doi       = {10.1016/B0-12-227410-5/00701-8},
  url       = {https://www.sciencedirect.com/science/article/pii/B0122274105007018}
}

@INPROCEEDINGS{2003ESASP.530..279S,
       author = {{Sch{\"u}ssler}, M. and {Sunrise Team}},
        title = "{Sunrise: balloon-borne high-resolution observation of the Sun}",
    booktitle = {European Rocket and Balloon Programmes and Related Research},
         year = 2003,
       editor = {{Warmbein}, Barbara},
       series = {ESA Special Publication},
       volume = {530},
        month = aug,
        pages = {279-283},
       adsurl = {https://ui.adsabs.harvard.edu/abs/2003ESASP.530..279S}
}

@article{article,
author = {Reiss, Martin and Temmer, M. and Rotter, T. and Hofmeister, Stefan and Veronig, Astrid},
year = {2014},
month = {08},
pages = {},
title = {Identification of coronal holes and filament channels in SDO/AIA 193\r{A} images via geometrical classification methods},
volume = {38},
journal = {Cent. Eur. Astrophys. Bull.}
}

@ARTICLE{2012SoPh..275...17L,
       author = {{Lemen}, James R. and {Title}, Alan M. and {Akin}, David J. and {Boerner}, Paul F. and {Chou}, Catherine and {Drake}, Jerry F. and {Duncan}, Dexter W. and {Edwards}, Christopher G. and {Friedlaender}, Frank M. and {Heyman}, Gary F. and {Hurlburt}, Neal E. and {Katz}, Noah L. and {Kushner}, Gary D. and {Levay}, Michael and {Lindgren}, Russell W. and {Mathur}, Dnyanesh P. and {McFeaters}, Edward L. and {Mitchell}, Sarah and {Rehse}, Roger A. and {Schrijver}, Carolus J. and {Springer}, Larry A. and {Stern}, Robert A. and {Tarbell}, Theodore D. and {Wuelser}, Jean-Pierre and {Wolfson}, C. Jacob and {Yanari}, Carl and {Bookbinder}, Jay A. and {Cheimets}, Peter N. and {Caldwell}, David and {Deluca}, Edward E. and {Gates}, Richard and {Golub}, Leon and {Park}, Sang and {Podgorski}, William A. and {Bush}, Rock I. and {Scherrer}, Philip H. and {Gummin}, Mark A. and {Smith}, Peter and {Auker}, Gary and {Jerram}, Paul and {Pool}, Peter and {Soufli}, Regina and {Windt}, David L. and {Beardsley}, Sarah and {Clapp}, Matthew and {Lang}, James and {Waltham}, Nicholas},
        title = "{The Atmospheric Imaging Assembly (AIA) on the Solar Dynamics Observatory (SDO)}",
      journal = {\solphys},
         year = 2012,
        month = jan,
       volume = {275},
       number = {1-2},
        pages = {17-40},
          doi = {10.1007/s11207-011-9776-8},
       adsurl = {https://ui.adsabs.harvard.edu/abs/2012SoPh..275...17L}
}

@ARTICLE{2012SoPh..275..207S,
       author = {{Scherrer}, P.~H. and {Schou}, J. and {Bush}, R.~I. and {Kosovichev}, A.~G. and {Bogart}, R.~S. and {Hoeksema}, J.~T. and {Liu}, Y. and {Duvall}, T.~L. and {Zhao}, J. and {Title}, A.~M. and {Schrijver}, C.~J. and {Tarbell}, T.~D. and {Tomczyk}, S.},
        title = "{The Helioseismic and Magnetic Imager (HMI) Investigation for the Solar Dynamics Observatory (SDO)}",
      journal = {\solphys},
         year = 2012,
        month = jan,
       volume = {275},
       number = {1-2},
        pages = {207-227},
          doi = {10.1007/s11207-011-9834-2},
       adsurl = {https://ui.adsabs.harvard.edu/abs/2012SoPh..275..207S}
}

@ARTICLE{2016SoPh..291.1887C,
       author = {{Couvidat}, S. and {Schou}, J. and {Hoeksema}, J.~T. and {Bogart}, R.~S. and {Bush}, R.~I. and {Duvall}, T.~L. and {Liu}, Y. and {Norton}, A.~A. and {Scherrer}, P.~H.},
        title = "{Observables Processing for the Helioseismic and Magnetic Imager Instrument on the Solar Dynamics Observatory}",
      journal = {\solphys},
         year = 2016,
        month = aug,
       volume = {291},
       number = {7},
        pages = {1887-1938},
          doi = {10.1007/s11207-016-0957-3},
archivePrefix = {arXiv},
       eprint = {1606.02368},
 primaryClass = {astro-ph.SR},
       adsurl = {https://ui.adsabs.harvard.edu/abs/2016SoPh..291.1887C}
}

@ARTICLE{2009SoPh..256...87K,
       author = {{Krista}, Larisza D. and {Gallagher}, Peter T.},
        title = "{Automated Coronal Hole Detection Using Local Intensity Thresholding Techniques}",
      journal = {\solphys},
         year = 2009,
        month = may,
       volume = {256},
       number = {1-2},
        pages = {87-100},
          doi = {10.1007/s11207-009-9357-2},
archivePrefix = {arXiv},
       eprint = {0905.1814},
 primaryClass = {astro-ph.SR},
       adsurl = {https://ui.adsabs.harvard.edu/abs/2009SoPh..256...87K}
}

@ARTICLE{sunpy_community2020,
  doi = {10.3847/1538-4357/ab4f7a},
  url = {https://iopscience.iop.org/article/10.3847/1538-4357/ab4f7a},
  author = {{The SunPy Community} and Barnes, Will T. and Bobra, Monica G. and Christe, Steven D. and Freij, Nabil and Hayes, Laura A. and Ireland, Jack and Mumford, Stuart and Perez-Suarez, David and Ryan, Daniel F. and Shih, Albert Y. and Chanda, Prateek and Glogowski, Kolja and Hewett, Russell and Hughitt, V. Keith and Hill, Andrew and Hiware, Kaustubh and Inglis, Andrew and Kirk, Michael S. F. and Konge, Sudarshan and Mason, James Paul and Maloney, Shane Anthony and Murray, Sophie A. and Panda, Asish and Park, Jongyeob and Pereira, Tiago M. D. and Reardon, Kevin and Savage, Sabrina and Sipőcz, Brigitta M. and Stansby, David and Jain, Yash and Taylor, Garrison and Yadav, Tannmay and Rajul and Dang, Trung Kien},
  title = {The SunPy Project: Open Source Development and Status of the Version 1.0 Core Package},
  journal = {The Astrophysical Journal},
  volume = {890},
  issue = {1},
  pages = {68-},
  publisher = {American Astronomical Society},
  year = {2020}
}

@ARTICLE{Mumford2020,
  doi = {10.21105/joss.01832},
  url = {https://doi.org/10.21105/joss.01832},
  year = {2020},
  publisher = {The Open Journal},
  volume = {5},
  number = {46},
  pages = {1832},
  author = {Stuart Mumford and Nabil Freij and Steven Christe and Jack Ireland and Florian Mayer and V. Hughitt and Albert Shih and Daniel Ryan and Simon Liedtke and David Pérez-Suárez and Pritish Chakraborty and Vishnunarayan K and Andrew Inglis and Punyaslok Pattnaik and Brigitta Sipőcz and Rishabh Sharma and Andrew Leonard and David Stansby and Russell Hewett and Alex Hamilton and Laura Hayes and Asish Panda and Matt Earnshaw and Nitin Choudhary and Ankit Kumar and Prateek Chanda and Md Haque and Michael Kirk and Michael Mueller and Sudarshan Konge and Rajul Srivastava and Yash Jain and Samuel Bennett and Ankit Baruah and Will Barnes and Michael Charlton and Shane Maloney and Nicky Chorley and Himanshu  and Sanskar Modi and James Mason and Naman9639  and Jose Rozo and Larry Manley and Agneet Chatterjee and John Evans and Michael Malocha and Monica Bobra and Sourav Ghosh and Airmansmith97  and Dominik Stańczak and Ruben De Visscher and Shresth Verma and Ankit Agrawal and Dumindu Buddhika and Swapnil Sharma and Jongyeob Park and Matt Bates and Dhruv Goel and Garrison Taylor and Goran Cetusic and Jacob  and Mateo Inchaurrandieta and Sally Dacie and Sanjeev Dubey and Deepankar Sharma and Erik Bray and Jai Rideout and Serge Zahniy and Tomas Meszaros and Abhigyan Bose and André Chicrala and Ankit  and Chloé Guennou and Daniel D'Avella and Daniel Williams and Jordan Ballew and Nick Murphy and Priyank Lodha and Thomas Robitaille and Yash Krishan and Andrew Hill and Arthur Eigenbrot and Benjamin Mampaey and Bernhard Wiedemann and Carlos Molina and Duygu Keşkek and Ishtyaq Habib and Joseph Letts and Juanjo Bazán and Quinn Arbolante and Reid Gomillion and Yash Kothari and Yash Sharma and Abigail Stevens and Adrian Price-Whelan and Ambar Mehrotra and Arseniy Kustov and Brandon Stone and Trung Dang and Emmanuel Arias and Fionnlagh Dover and Freek Verstringe and Gulshan Kumar and Harsh Mathur and Igor Babuschkin and Jaylen Wimbish and Juan Buitrago-Casas and Kalpesh Krishna and Kaustubh Hiware and Manas Mangaonkar and Matthew Mendero and Mickaël Schoentgen and Norbert Gyenge and Ole Streicher and Rajasekhar Mekala and Rishabh Mishra and Shashank Srikanth and Sarthak Jain and Tannmay Yadav and Tessa Wilkinson and Tiago Pereira and Yudhik Agrawal and Jamescalixto  and Yasintoda  and Sophie Murray},
  title = {SunPy: A Python package for Solar Physics},
  journal = {Journal of Open Source Software}
}

@ARTICLE{2023RLSFN..34.1045C,
       author = {{Cantoresi}, Matteo and {Berrilli}, Francesco and {Lepreti}, Fabio},
        title = "{Organization scale of photospheric magnetic imbalance in coronal holes}",
      journal = {Rendiconti Lincei. Scienze Fisiche e Naturali},
         year = 2023,
        month = dec,
       volume = {34},
       number = {4},
        pages = {1045-1053},
          doi = {10.1007/s12210-023-01185-x},
       adsurl = {https://ui.adsabs.harvard.edu/abs/2023RLSFN..34.1045C}
}

@ARTICLE{2009SoPh..260...43A,
       author = {{Abramenko}, Valentyna and {Yurchyshyn}, Vasyl and {Watanabe}, Hiroko},
        title = "{Parameters of the Magnetic Flux inside Coronal Holes}",
      journal = {\solphys},
         year = 2009,
        month = nov,
       volume = {260},
       number = {1},
        pages = {43-57},
          doi = {10.1007/s11207-009-9433-7},
archivePrefix = {arXiv},
       eprint = {0908.2460},
 primaryClass = {astro-ph.SR},
       adsurl = {https://ui.adsabs.harvard.edu/abs/2009SoPh..260...43A}
}

@ARTICLE{1979SSRv...23..139H,
       author = {{Harvey}, J.~W. and {Sheeley}, N.~R., Jr.},
        title = "{Coronal Holes and Solar Magnetic Fields (Article published in the special issues: Proceedings of the Symposium on Solar Terrestrial Physics held in Innsbruck, May- June 1978. (pp. 137-538))}",
      journal = {\ssr},
         year = 1979,
        month = apr,
       volume = {23},
       number = {2},
        pages = {139-158},
          doi = {10.1007/BF00173808},
       adsurl = {https://ui.adsabs.harvard.edu/abs/1979SSRv...23..139H}
}

@ARTICLE{2021ApJ...913...28R,
       author = {{Reiss}, Martin A. and {Muglach}, Karin and {M{\"o}stl}, Christian and {Arge}, Charles N. and {Bailey}, Rachel and {Delouille}, V{\'e}ronique and {Garton}, Tadhg M. and {Hamada}, Amr and {Hofmeister}, Stefan and {Illarionov}, Egor and {Jarolim}, Robert and {Kirk}, Michael S.~F. and {Kosovichev}, Alexander and {Krista}, Larisza and {Lee}, Sangwoo and {Lowder}, Chris and {MacNeice}, Peter J. and {Veronig}, Astrid and {Cospar Iswat Coronal Hole Boundary Working Team}},
        title = "{The Observational Uncertainty of Coronal Hole Boundaries in Automated Detection Schemes}",
      journal = {\apj},
         year = 2021,
        month = may,
       volume = {913},
       number = {1},
          eid = {28},
        pages = {28},
          doi = {10.3847/1538-4357/abf2c8},
archivePrefix = {arXiv},
       eprint = {2103.14403},
 primaryClass = {astro-ph.SR},
       adsurl = {https://ui.adsabs.harvard.edu/abs/2021ApJ...913...28R}
}

@ARTICLE{2019SoPh..294..144H,
       author = {{Heinemann}, Stephan G. and {Temmer}, Manuela and {Heinemann}, Niko and {Dissauer}, Karin and {Samara}, Evangelia and {Jer{\v{c}}i{\'c}}, Veronika and {Hofmeister}, Stefan J. and {Veronig}, Astrid M.},
        title = "{Statistical Analysis and Catalog of Non-polar Coronal Holes Covering the SDO-Era Using CATCH}",
      journal = {\solphys},
         year = 2019,
        month = oct,
       volume = {294},
       number = {10},
          eid = {144},
        pages = {144},
          doi = {10.1007/s11207-019-1539-y},
archivePrefix = {arXiv},
       eprint = {1907.01990},
 primaryClass = {astro-ph.SR},
       adsurl = {https://ui.adsabs.harvard.edu/abs/2019SoPh..294..144H}
}

@article{Reiss_2021,
doi = {10.3847/1538-4357/abf2c8},
url = {https://dx.doi.org/10.3847/1538-4357/abf2c8},
year = {2021},
month = {may},
publisher = {The American Astronomical Society},
volume = {913},
number = {1},
pages = {28},
author = {Martin A. Reiss and Karin Muglach and Christian Möstl and Charles N. Arge and Rachel Bailey and Véronique Delouille and Tadhg M. Garton and Amr Hamada and Stefan Hofmeister and Egor Illarionov and Robert Jarolim and Michael S. F. Kirk and Alexander Kosovichev and Larisza Krista and Sangwoo Lee and Chris Lowder and Peter J. MacNeice and Astrid Veronig and COSPAR ISWAT Coronal Hole Boundary Working Team},
title = {The Observational Uncertainty of Coronal Hole Boundaries in Automated Detection Schemes},
journal = {The Astrophysical Journal}
}

@ARTICLE{2004SoPh..225..227W,
       author = {{Wiegelmann}, T. and {Solanki}, S.~K.},
        title = "{Similarities and Differences between Coronal Holes and the Quiet Sun: Are Loop Statistics the Key?}",
      journal = {\solphys},
         year = 2004,
        month = dec,
       volume = {225},
       number = {2},
        pages = {227-247},
          doi = {10.1007/s11207-004-3747-2},
archivePrefix = {arXiv},
       eprint = {0802.0120},
 primaryClass = {astro-ph},
       adsurl = {https://ui.adsabs.harvard.edu/abs/2004SoPh..225..227W}
}

@ARTICLE{2018JGRA..123.1738H,
       author = {{Hofmeister}, Stefan J. and {Veronig}, Astrid and {Temmer}, Manuela and {Vennerstrom}, Susanne and {Heber}, Bernd and {Vr{\v{s}}nak}, Bojan},
        title = "{The Dependence of the Peak Velocity of High-Speed Solar Wind Streams as Measured in the Ecliptic by ACE and the STEREO satellites on the Area and Co-latitude of Their Solar Source Coronal Holes}",
      journal = {Journal of Geophysical Research (Space Physics)},
         year = 2018,
        month = mar,
       volume = {123},
       number = {3},
        pages = {1738-1753},
          doi = {10.1002/2017JA024586},
archivePrefix = {arXiv},
       eprint = {1804.09579},
 primaryClass = {astro-ph.SR},
       adsurl = {https://ui.adsabs.harvard.edu/abs/2018JGRA..123.1738H}
}

@ARTICLE{2012SoPh..275....3P,
       author = {{Pesnell}, W. Dean and {Thompson}, B.~J. and {Chamberlin}, P.~C.},
        title = "{The Solar Dynamics Observatory (SDO)}",
      journal = {\solphys},
         year = 2012,
        month = jan,
       volume = {275},
       number = {1-2},
        pages = {3-15},
          doi = {10.1007/s11207-011-9841-3},
       adsurl = {https://ui.adsabs.harvard.edu/abs/2012SoPh..275....3P}
}

@ARTICLE{2012SoPh..275..229S,
       author = {{Schou}, J. and {Scherrer}, P.~H. and {Bush}, R.~I. and {Wachter}, R. and {Couvidat}, S. and {Rabello-Soares}, M.~C. and {Bogart}, R.~S. and {Hoeksema}, J.~T. and {Liu}, Y. and {Duvall}, T.~L. and {Akin}, D.~J. and {Allard}, B.~A. and {Miles}, J.~W. and {Rairden}, R. and {Shine}, R.~A. and {Tarbell}, T.~D. and {Title}, A.~M. and {Wolfson}, C.~J. and {Elmore}, D.~F. and {Norton}, A.~A. and {Tomczyk}, S.},
        title = "{Design and Ground Calibration of the Helioseismic and Magnetic Imager (HMI) Instrument on the Solar Dynamics Observatory (SDO)}",
      journal = {\solphys},
         year = 2012,
        month = jan,
       volume = {275},
       number = {1-2},
        pages = {229-259},
          doi = {10.1007/s11207-011-9842-2},
       adsurl = {https://ui.adsabs.harvard.edu/abs/2012SoPh..275..229S}
}

@ARTICLE{2023A&A...678A.196N,
       author = {{N{\"o}lke}, J.~D. and {Solanki}, S.~K. and {Hirzberger}, J. and {Peter}, H. and {Chitta}, L.~P. and {Kahil}, F. and {Valori}, G. and {Wiegelmann}, T. and {Orozco Su{\'a}rez}, D. and {Albert}, K. and {Albelo Jorge}, N. and {Appourchaux}, T. and {Alvarez-Herrero}, A. and {Blanco Rodr{\'\i}guez}, J. and {Gandorfer}, A. and {Germerott}, D. and {Guerrero}, L. and {Gutierrez-Marques}, P. and {Kolleck}, M. and {del Toro Iniesta}, J.~C. and {Volkmer}, R. and {Woch}, J. and {Fiethe}, B. and {G{\'o}mez Cama}, J.~M. and {P{\'e}rez-Grande}, I. and {Sanchis Kilders}, E. and {Balaguer Jim{\'e}nez}, M. and {Bellot Rubio}, L.~R. and {Calchetti}, D. and {Carmona}, M. and {Deutsch}, W. and {Feller}, A. and {Fernandez-Rico}, G. and {Fern{\'a}ndez-Medina}, A. and {Garc{\'\i}a Parejo}, P. and {Gasent Blesa}, J.~L. and {Gizon}, L. and {Grauf}, B. and {Heerlein}, K. and {Korpi-Lagg}, A. and {Lange}, T. and {L{\'o}pez Jim{\'e}nez}, A. and {Maue}, T. and {Meller}, R. and {Moreno Vacas}, A. and {M{\"u}ller}, R. and {Nakai}, E. and {Schmidt}, W. and {Schou}, J. and {Sch{\"u}hle}, U. and {Sinjan}, J. and {Staub}, J. and {Strecker}, H. and {Torralbo}, I. and {Berghmans}, D. and {Kraaikamp}, E. and {Rodriguez}, L. and {Verbeeck}, C. and {Zhukov}, A.~N. and {Auchere}, F. and {Buchlin}, E. and {Parenti}, S. and {Janvier}, M. and {Barczynski}, K. and {Harra}, L. and {Schwanitz}, C. and {Aznar Cuadrado}, R. and {Mandal}, S. and {Teriaca}, L. and {Long}, D. and {Smith}, P.},
        title = "{Coronal voids and their magnetic nature}",
      journal = {\aap},
         year = 2023,
        month = oct,
       volume = {678},
          eid = {A196},
        pages = {A196},
          doi = {10.1051/0004-6361/202346040},
archivePrefix = {arXiv},
       eprint = {2309.09789},
 primaryClass = {astro-ph.SR},
       adsurl = {https://ui.adsabs.harvard.edu/abs/2023A&A...678A.196N}
}

@ARTICLE{2014CEAB...38...95R,
       author = {{Reiss}, M. and {Temmer}, M. and {Rotter}, T. and {Hofmeister}, S.~J. and {Veronig}, A.~M.},
        title = "{Identification of coronal holes and filament channels in SDO/AIA 193{\r{A}} images via geometrical classification methods}",
      journal = {Central European Astrophysical Bulletin},
         year = 2014,
        month = jan,
       volume = {38},
        pages = {95-104},
          doi = {10.48550/arXiv.1408.2777},
archivePrefix = {arXiv},
       eprint = {1408.2777},
 primaryClass = {astro-ph.SR},
       adsurl = {https://ui.adsabs.harvard.edu/abs/2014CEAB...38...95R}
}

@ARTICLE{2006SoPh..239..337J,
       author = {{Jian}, L. and {Russell}, C.~T. and {Luhmann}, J.~G. and {Skoug}, R.~M.},
        title = "{Properties of Stream Interactions at One AU During 1995   2004}",
      journal = {\solphys},
         year = 2006,
        month = dec,
       volume = {239},
       number = {1-2},
        pages = {337-392},
          doi = {10.1007/s11207-006-0132-3},
       adsurl = {https://ui.adsabs.harvard.edu/abs/2006SoPh..239..337J}
}

@ARTICLE{2009SoPh..259..345J,
       author = {{Jian}, L.~K. and {Russell}, C.~T. and {Luhmann}, J.~G. and {Galvin}, A.~B. and {MacNeice}, P.~J.},
        title = "{Multi-Spacecraft Observations: Stream Interactions and Associated Structures}",
      journal = {\solphys},
         year = 2009,
        month = oct,
       volume = {259},
       number = {1-2},
          eid = {345},
        pages = {345},
          doi = {10.1007/s11207-009-9445-3},
       adsurl = {https://ui.adsabs.harvard.edu/abs/2009SoPh..259..345J}
}

@ARTICLE{2018LRSP...15....1R,
       author = {{Richardson}, Ian G.},
        title = "{Solar wind stream interaction regions throughout the heliosphere}",
      journal = {Living Reviews in Solar Physics},
         year = 2018,
        month = dec,
       volume = {15},
       number = {1},
          eid = {1},
        pages = {1},
          doi = {10.1007/s41116-017-0011-z},
       adsurl = {https://ui.adsabs.harvard.edu/abs/2018LRSP...15....1R}
}

@online{pyspedas_doc,
  author       = {{Space Environment Modeling Group}},
  title        = {{PySPEDAS Documentation}},
  year         = {2020},
  url          = {https://pyspedas.readthedocs.io/en/latest/},
  urldate      = {2025-07-11}
}

@ARTICLE{2019A&A...629A..22H,
       author = {{Hofmeister}, Stefan J. and {Utz}, Dominik and {Heinemann}, Stephan G. and {Veronig}, Astrid and {Temmer}, Manuela},
        title = "{Photospheric magnetic structure of coronal holes}",
      journal = {\aap},
         year = 2019,
        month = sep,
       volume = {629},
          eid = {A22},
        pages = {A22},
          doi = {10.1051/0004-6361/201935918},
archivePrefix = {arXiv},
       eprint = {1909.03806},
 primaryClass = {astro-ph.SR},
       adsurl = {https://ui.adsabs.harvard.edu/abs/2019A&A...629A..22H}
}

@ARTICLE{2017ApJ...835..268H,
       author = {{Hofmeister}, Stefan J. and {Veronig}, Astrid and {Reiss}, Martin A. and {Temmer}, Manuela and {Vennerstrom}, Susanne and {Vr{\v{s}}nak}, Bojan and {Heber}, Bernd},
        title = "{Characteristics of Low-latitude Coronal Holes near the Maximum of Solar Cycle 24}",
      journal = {\apj},
         year = 2017,
        month = feb,
       volume = {835},
       number = {2},
          eid = {268},
        pages = {268},
          doi = {10.3847/1538-4357/835/2/268},
archivePrefix = {arXiv},
       eprint = {1702.02050},
 primaryClass = {astro-ph.SR},
       adsurl = {https://ui.adsabs.harvard.edu/abs/2017ApJ...835..268H}
}

@ARTICLE{2019JGRA..124.3871G,
       author = {{Grandin}, Maxime and {Aikio}, Anita T. and {Kozlovsky}, Alexander},
        title = "{Properties and Geoeffectiveness of Solar Wind High-Speed Streams and Stream Interaction Regions During Solar Cycles 23 and 24}",
      journal = {Journal of Geophysical Research (Space Physics)},
         year = 2019,
        month = jun,
       volume = {124},
       number = {6},
        pages = {3871-3892},
          doi = {10.1029/2018JA026396},
archivePrefix = {arXiv},
       eprint = {2006.06302},
 primaryClass = {physics.space-ph},
       adsurl = {https://ui.adsabs.harvard.edu/abs/2019JGRA..124.3871G}
}

@ARTICLE{2012SoPh..281..793R,
       author = {{Rotter}, T. and {Veronig}, A.~M. and {Temmer}, M. and {Vr{\v{s}}nak}, B.},
        title = "{Relation Between Coronal Hole Areas on the Sun and the Solar Wind Parameters at 1 AU}",
      journal = {\solphys},
         year = 2012,
        month = dec,
       volume = {281},
       number = {2},
        pages = {793-813},
          doi = {10.1007/s11207-012-0101-y},
       adsurl = {https://ui.adsabs.harvard.edu/abs/2012SoPh..281..793R}
}

@ARTICLE{2014A&A...561A..29V,
       author = {{Verbeeck}, C. and {Delouille}, V. and {Mampaey}, B. and {De Visscher}, R.},
        title = "{The SPoCA-suite: Software for extraction, characterization, and tracking of active regions and coronal holes on EUV images}",
      journal = {\aap},
         year = 2014,
        month = jan,
       volume = {561},
          eid = {A29},
        pages = {A29},
          doi = {10.1051/0004-6361/201321243},
       adsurl = {https://ui.adsabs.harvard.edu/abs/2014A&A...561A..29V}
}

@ARTICLE{2018JSWSC...8A...2G,
       author = {{Garton}, Tadhg M. and {Gallagher}, Peter T. and {Murray}, Sophie A.},
        title = "{Automated coronal hole identification via multi-thermal intensity segmentation}",
      journal = {Journal of Space Weather and Space Climate},
         year = 2018,
        month = jan,
       volume = {8},
          eid = {A02},
        pages = {A02},
          doi = {10.1051/swsc/2017039},
archivePrefix = {arXiv},
       eprint = {1711.11476},
 primaryClass = {astro-ph.SR},
       adsurl = {https://ui.adsabs.harvard.edu/abs/2018JSWSC...8A...2G}
}

@ARTICLE{2016ApJ...823...53C,
       author = {{Caplan}, R.~M. and {Downs}, C. and {Linker}, J.~A.},
        title = "{Synchronic Coronal Hole Mapping Using Multi-instrument EUV Images: Data Preparation and Detection Method}",
      journal = {\apj},
         year = 2016,
        month = may,
       volume = {823},
       number = {1},
          eid = {53},
        pages = {53},
          doi = {10.3847/0004-637X/823/1/53},
archivePrefix = {arXiv},
       eprint = {1510.04718},
 primaryClass = {astro-ph.SR},
       adsurl = {https://ui.adsabs.harvard.edu/abs/2016ApJ...823...53C}
}

@ARTICLE{2021ApJ...918...21L,
       author = {{Linker}, Jon A. and {Heinemann}, Stephan G. and {Temmer}, Manuela and {Owens}, Mathew J. and {Caplan}, Ronald M. and {Arge}, Charles N. and {Asvestari}, Eleanna and {Delouille}, Veronique and {Downs}, Cooper and {Hofmeister}, Stefan J. and {Jebaraj}, Immanuel C. and {Madjarska}, Maria S. and {Pinto}, Rui F. and {Pomoell}, Jens and {Samara}, Evangelia and {Scolini}, Camilla and {Vr{\v{s}}nak}, Bojan},
        title = "{Coronal Hole Detection and Open Magnetic Flux}",
      journal = {\apj},
         year = 2021,
        month = sep,
       volume = {918},
       number = {1},
          eid = {21},
        pages = {21},
          doi = {10.3847/1538-4357/ac090a},
archivePrefix = {arXiv},
       eprint = {2103.05837},
 primaryClass = {astro-ph.SR},
       adsurl = {https://ui.adsabs.harvard.edu/abs/2021ApJ...918...21L}
}

@ARTICLE{2005JGRA..110.4104S,
       author = {{Schwadron}, N.~A. and {McComas}, D.~J. and {Elliott}, H.~A. and {Gloeckler}, G. and {Geiss}, J. and {von Steiger}, R.},
        title = "{Solar wind from the coronal hole boundaries}",
      journal = {Journal of Geophysical Research (Space Physics)},
         year = 2005,
        month = apr,
       volume = {110},
       number = {A4},
          eid = {A04104},
        pages = {A04104},
          doi = {10.1029/2004JA010896},
       adsurl = {https://ui.adsabs.harvard.edu/abs/2005JGRA..110.4104S}
}

@ARTICLE{2022ApJ...931...96A,
       author = {{Aslanyan}, V. and {Pontin}, D.~I. and {Scott}, R.~B. and {Higginson}, A.~K. and {Wyper}, P.~F. and {Antiochos}, S.~K.},
        title = "{The Dynamic Structure of Coronal Hole Boundaries}",
      journal = {\apj},
         year = 2022,
        month = jun,
       volume = {931},
       number = {2},
          eid = {96},
        pages = {96},
          doi = {10.3847/1538-4357/ac69ed},
       adsurl = {https://ui.adsabs.harvard.edu/abs/2022ApJ...931...96A}
}

@ARTICLE{2016SoPh..291.2329B,
       author = {{Bilenko}, Irina A. and {Tavastsherna}, Ksenia S.},
        title = "{Coronal Hole and Solar Global Magnetic Field Evolution in 1976 - 2012}",
      journal = {\solphys},
         year = 2016,
        month = oct,
       volume = {291},
       number = {8},
        pages = {2329-2352},
          doi = {10.1007/s11207-016-0966-2},
archivePrefix = {arXiv},
       eprint = {1805.09543},
 primaryClass = {astro-ph.SR},
       adsurl = {https://ui.adsabs.harvard.edu/abs/2016SoPh..291.2329B}
}

@ARTICLE{2017SoPh..292...18L,
       author = {{Lowder}, Chris and {Qiu}, Jiong and {Leamon}, Robert},
        title = "{Coronal Holes and Open Magnetic Flux over Cycles 23 and 24}",
      journal = {\solphys},
         year = 2017,
        month = jan,
       volume = {292},
       number = {1},
          eid = {18},
        pages = {18},
          doi = {10.1007/s11207-016-1041-8},
archivePrefix = {arXiv},
       eprint = {1612.07595},
 primaryClass = {astro-ph.SR},
       adsurl = {https://ui.adsabs.harvard.edu/abs/2017SoPh..292...18L}
}

@misc{katuwal2025,
  author       = {Katuwal, Khagendra},
  title        = {Supplementary Materials for ``The Unipolarity of Magnetic Field in the Equatorial Coronal Holes''},
  year         = {2025},
  version      = {1.0.0},
  publisher    = {Zenodo},
  doi          = {10.5281/zenodo.16521356},
  url          = {https://doi.org/10.5281/zenodo.16521356},
  note         = {Data set}
}

@ARTICLE{2012A&A...548A..62H,
       author = {{Huang}, Z. and {Madjarska}, M.~S. and {Doyle}, J.~G. and {Lamb}, D.~A.},
        title = "{Coronal hole boundaries at small scales. IV. SOT view. Magnetic field properties of small-scale transient brightenings in coronal holes}",
      journal = {\aap},
         year = 2012,
        month = dec,
       volume = {548},
          eid = {A62},
        pages = {A62},
          doi = {10.1051/0004-6361/201220079},
archivePrefix = {arXiv},
       eprint = {1210.2009},
 primaryClass = {astro-ph.SR},
       adsurl = {https://ui.adsabs.harvard.edu/abs/2012A&A...548A..62H}
}

@ARTICLE{2010ApJ...719..131I,
       author = {{Ito}, Hiroaki and {Tsuneta}, Saku and {Shiota}, Daikou and {Tokumaru}, Munetoshi and {Fujiki}, Ken'ichi},
        title = "{Is the Polar Region Different from the Quiet Region of the Sun?}",
      journal = {\apj},
         year = 2010,
        month = aug,
       volume = {719},
       number = {1},
        pages = {131-142},
          doi = {10.1088/0004-637X/719/1/131},
archivePrefix = {arXiv},
       eprint = {1005.3667},
 primaryClass = {astro-ph.SR},
       adsurl = {https://ui.adsabs.harvard.edu/abs/2010ApJ...719..131I}
}

@ARTICLE{2024ApJS..271....6R,
       author = {{Reiss}, Martin A. and {Muglach}, Karin and {Mason}, Emily and {Davies}, Emma E. and {Chakraborty}, Shibaji and {Delouille}, Veronique and {Downs}, Cooper and {Garton}, Tadhg M. and {Grajeda}, Jeremy A. and {Hamada}, Amr and {Heinemann}, Stephan G. and {Hofmeister}, Stefan and {Illarionov}, Egor and {Jarolim}, Robert and {Krista}, Larisza and {Lowder}, Chris and {Verwichte}, Erwin and {Arge}, Charles N. and {Boucheron}, Laura E. and {Foullon}, Claire and {Kirk}, Michael S. and {Kosovichev}, Alexander and {Leisner}, Andrew and {M{\"o}stl}, Christian and {Turtle}, James and {Veronig}, Astrid},
        title = "{A Community Data Set for Comparing Automated Coronal Hole Detection Schemes}",
      journal = {\apjs},
         year = 2024,
        month = mar,
       volume = {271},
       number = {1},
          eid = {6},
        pages = {6},
          doi = {10.3847/1538-4365/ad1408},
       adsurl = {https://ui.adsabs.harvard.edu/abs/2024ApJS..271....6R}
}

@ARTICLE{2014SoPh..289.3483H,
       author = {{Hoeksema}, J. Todd and {Liu}, Yang and {Hayashi}, Keiji and {Sun}, Xudong and {Schou}, Jesper and {Couvidat}, Sebastien and {Norton}, Aimee and {Bobra}, Monica and {Centeno}, Rebecca and {Leka}, K.~D. and {Barnes}, Graham and {Turmon}, Michael},
        title = "{The Helioseismic and Magnetic Imager (HMI) Vector Magnetic Field Pipeline: Overview and Performance}",
      journal = {\solphys},
         year = 2014,
        month = sep,
       volume = {289},
       number = {9},
        pages = {3483-3530},
          doi = {10.1007/s11207-014-0516-8},
archivePrefix = {arXiv},
       eprint = {1404.1881},
 primaryClass = {astro-ph.SR},
       adsurl = {https://ui.adsabs.harvard.edu/abs/2014SoPh..289.3483H}
}

@article{mostl2020prediction,
  title={Prediction of the in situ coronal mass ejection rate for solar cycle 25: Implications for parker solar probe in situ observations},
  author={M{\"o}stl, Christian and Weiss, Andreas J and Bailey, Rachel L and Reiss, Martin A and Amerstorfer, Tanja and Hinterreiter, J{\"u}rgen and Bauer, Maike and McIntosh, Scott W and Lugaz, No{\'e} and Stansby, David},
  journal={The Astrophysical Journal},
  volume={903},
  number={2},
  pages={92},
  year={2020},
  publisher={IOP Publishing}
}

@article{silwal2024multispecies,
  title={Multispecies Energetic Particle Acceleration Associated with CIR and ICME-driven Shocks},
  author={Silwal, Ashok and Zhao, Lingling and Zank, Gary P and Wang, Bingbing and Pit{\~n}a, Alexander and Gautam, Sujan Prasad and Park, Byeongseon and Nakanotani, Masaru and Zhu, Xingyu},
  journal={The Astrophysical Journal},
  volume={972},
  number={2},
  pages={168},
  year={2024},
  publisher={IOP Publishing}
}

\end{document}